\documentclass[11pt,oneside,letterpaper]{article}
\usepackage{amssymb}
\usepackage{amsmath}
\usepackage[dvips]{graphicx}
\usepackage{setspace}
\usepackage{fancyhdr}
\usepackage{xcolor}
\usepackage{amsmath}

\usepackage{ifpdf}
\usepackage{graphicx,wrapfig,lipsum}
\usepackage{rotating}
\usepackage{comment}
\usepackage[colorlinks=true,citecolor=darkgreen,linkcolor=purple,urlcolor=purple]{hyperref}

\DeclareMathOperator\arctanh{arctanh}

\definecolor{darkgreen}{rgb}{0,0.5,0}
\definecolor{darkblue}{rgb}{0,0,0.6}
\definecolor{purple}{rgb}{0.4,.2,0.7}
\definecolor{awesome}{rgb}{1.0, 0.13, 0.32}

\newcommand{\bi}{\begin{itemize}}
\newcommand{\ei}{\end{itemize}}
\newcommand{\bea}{\begin{eqnarray}}
\newcommand{\eea}{\end{eqnarray}}
\newcommand{\be}{\begin{equation}}
\newcommand{\ee}{\end{equation}}

 \newcommand{\red}[1]{{\color{red}#1}}

\numberwithin{equation}{section}
\allowdisplaybreaks  % allow page breaks in displayed eqs

\begin{document}

\vspace*{1.5cm}
\begin{center}
{ \LARGE \textsc{Infrared Realization of dS$_2$ in AdS$_2$}}% holography}}
\\ \vspace*{1.7cm}
Dionysios Anninos$^1$ and Diego M. Hofman$^2$
%Circle saddles and Centaur geometries }}%Circle Saddles and Ultraviolet Merging Remarks on the entropy of De Sitter space Revisited}\\}
\\
\vspace*{1.7cm}
$^1$ Institute for Advanced Study, 1 Einstein Drive, Princeton, NJ 85040 \\
$^2$ Institute for Theoretical Physics Amsterdam and Delta Institute for Theoretical Physics, University of Amsterdam, Science Park 904, 1098 XH Amsterdam, The Netherlands 
\vspace*{0.6cm}
%\newline

%\vspace*{0.6cm}
%\newline

%\vspace*{0.6cm}

%\vspace*{0.8cm}

\end{center}
\vspace*{1.5cm}
\begin{abstract}
\noindent

We describe a two-dimensional geometry that smoothly interpolates between an asymptotically AdS$_2$ geometry and the static patch of dS$_2$. We find this `centaur' geometry to be a solution of dilaton gravity with a specific class of potentials for the dilaton. We interpret the centaur geometry as a thermal state in the putative quantum mechanics dual to the AdS$_2$ evolved with the global Hamiltonian. We compute the thermodynamic properties and observe that the centaur state has finite entropy and positive specific heat. The static patch is the infrared part of the centaur geometry. We discuss boundary observables sensitive to the static patch region.%Circle saddles in simple models.
%We squash the living bejesus out of AdS$_3$ and find black holes with the living bejesus squashed out of them.

\end{abstract}

\newpage
\setcounter{page}{1}
\pagenumbering{arabic}

%\tableofcontents
\setcounter{tocdepth}{2}

%\cfoot{\thepage}
\onehalfspacing

\section{Introduction}

We have come to think of theories of gravity containing spacetimes endowed with an asymptotic spatial boundary as quantum mechanical systems with a large number of degrees of freedom. The most concrete examples involve spacetimes which asymptote to a $(d+1)$-dimensional anti-de Sitter geometry \cite{Maldacena:1997re}, in which case the quantum system in the ultraviolet is a $d$-dimensional quantum field theory enjoying conformal symmetries.  In the special case $d=1$, the theory is a conformally invariant quantum mechanics describing an AdS$_2$ geometry, such as the one appearing in the near horizon of black holes with a vanishing horizon temperature.  Though such systems are more delicate than their quantum field theoretic counterparts, there have been recent attempts to materialize this important case \cite{Sachdev:1992fk,pg,kitaev,Sachdev:2015efa,Anninos:2013nra,Anninos:2016szt,Polchinski:2016xgd,Maldacena:2016hyu,Maldacena:2016upp}. Upon turning on relevant deformations, quantum systems can flow away from the fixed point to the infrared. Correspondingly, the deep interior of the dual geometry can be deformed away from the exact anti-de Sitter geometry and can manifest a host of interesting infrared physics, such as the emergence of horizons \cite{Hartnoll:2009sz} or even the capping off of space altogether \cite{Polchinski:2000uf}. This remains the case, even for asymptotically AdS$_2$ geometries \cite{Majumdar:1947eu,papapetrou}. 

In this paper, we investigate a geometry that is asymptotically AdS$_2$ but flows in the interior to the static patch of a two-dimensional de Sitter geometry. We refer to such geometries as centaur geometries. The static patch of a de Sitter spacetime is the intersection of the future and past causal diamonds of a de Sitter observer. In the current cosmological era, which is dominated by a positive cosmological constant, we find ourselves to be well described by such static patch observers. Thus, the holography of a static patch geometry is a pertinent matter \cite{Parikh:2004wh,Gibbons:1977mu,Banks:2012ic,Dong:2010pm,Goheer:2002vf,Anninos:2012qw,Verlinde:2016toy,Maltz:2016max}. On its own, the static patch does not enjoy the crucial feature of an asymptotic spatial boundary, thus making the question a hard one. By constructing centaur geometries, we are able to formulate questions about the static patch interior and interpret them in terms of the putative quantum mechanical dual of the asymptotically AdS$_2$ geometry. The centaur state is a particular excited state. Since it has a horizon, we may further interpret it as a thermal state.

In what follows we explore several aspects of the centaur geometry. We begin with some general remarks about geometries holographically dual to quantum mechanical systems. By this we mean $(0+1)$-dimensional quantum systems with no spatial locality. We stress the fact that the UV behavior of quantum mechanics is universal, as opposed to the more familiar situation in higher dimensional quantum field theory. Then, we identify the centaur geometry as a solution to a particular class of dilaton gravity theories and  study its thermodynamic properties, and the properties of waves propagating in the centaur geometries. Long lived particles travelling along the static patch inertial worldline manifest themselves in the pole structure of the boundary retarded Green function. We conclude with speculative remarks on the higher dimensional case, and possible microscopic constructions.
%and the relevance to string theory.  

Throughout this work we consider only the two dimensional example of dS$_2$ infrared geometries completed in the ultraviolet by AdS$_2$. This is just a matter of simplicity. We expect that the same construction can be generalized to higher dimensions by embedding dS$_d$ in AdS$_2 \times S^{d-2}$ ultraviolet geometries. Notice that the celestial sphere in this geometry flows monotonically, increasing in size as it reaches the cosmological horizon. This feature is absent in previous attempts to embed dS$_d$ in AdS$_d$ geometries \cite{Freivogel:2005qh,Lowe:2010np}. We return to this issue in the discussion.

\section{UV behavior of QM vs. worldline holography}\label{UVQM}

In this section we make some general remarks regarding properties of the geometry dual to a quantum mechanical theory. Unlike quantum field theory, the space of relevant deformations is very rich in quantum mechanics and is hard to make any sensible classification of infrared behaviors. On the other hand, the high energy behavior of quantum mechanics is rather constrained, and we consider here several different cases and the corresponding geometric features.

\subsection{`Infinite' quantum mechanics}

Consider first a theory containing bosons $x_I$ and fermions $\psi_i$, with canonical kinetic terms and smooth potential:
\begin{equation}
S = \int dt \left[ \dot{x}_{I} \dot{x}_{I} + i \, \bar{\psi}_{i} \dot{\psi}_i - V\left( x_I, \bar{\psi}_i, \psi_i \right) \right]~,
\end{equation}
where $(I,i)$ denote a general index structure, i.e. they could be, for example, matrix or vector indices. At high energies, compared to any other scale in the problem, energy eigenstates are well approximated by oscillatory functions $\Psi_E(x_I) \sim e^{i {k}_I x_I}$, with ${k}_I {k}_I = E$.
The form of the wavefunction does {not} depend on the form of the details of the potential and is thus universal. In Wilsonian terms, there are no irrelevant deformations in $(0+1)$-dimensional field theory with canonical kinetic terms. If no singularities in the potential are allowed, the theory is free in the UV (which in the absence of a spatial structure is the high energy regime). If singularities are allowed, there exist marginal deformations, which we discuss in the next subsection. 

A holographic construction of this setup appears manifestly in the worldline theory of a stack of $N$ D0-branes dual the D0-brane near horizon geometry \cite{Banks:1996vh,Itzhaki:1998dd}. Due to the radial dependence of the dilaton, the eight-sphere shrinks indefinitely in the string frame. The string frame metric and dilaton are given by:
\begin{equation}
\frac{ds^2}{\alpha'} = -\frac{U^{7/2}}{\sqrt{a \lambda}} dt^2 + {\sqrt{a \lambda}} \left( \frac{dU^2}{U^{7/2}} + \frac{d\Omega_8^2}{U^{3/2}} \right)~, \quad e^\Phi =  \frac{1}{N} \, (2\pi)^2 \, a^{3/4} \left(\frac{U}{\lambda^{1/3}}\right)^{-21/4}~,
\end{equation}
%(where distances are measured with respect to the string length rather than the Planck length) 
%We have defined:
%\begin{equation}
%a = 240 \, \pi^5~, \quad\quad f(U) = 1-(U_0/U)^{7}~.
%\end{equation}
with $a = 240 \, \pi^5$ and $\lambda$ the 't Hooft coupling of the D0-brane matrix quantum mechanics. The supergravity approximation breaks down at sufficiently large $U$, and must be replaced with a string field theory.
%Thus, an observer that remains large with respect to the string scale will perceive the physics near the asymptotic sphere as stringy. 
The Einstein frame metric is given by
$g^{(Einstein)}_{\mu\nu} = e^{-\Phi/2} g^{(string)}_{\mu\nu}$.
%\end{equation}
%Introducing a radial variable $R$ for which grows the sphere grows as $R^2 d\Omega^2_8$ asymptotically, we have:
%\begin{equation}
%g_{tt}^{(Einstein)} \sim R^{98/9}~.
%\end{equation}
In the Einstein frame both the $g_{tt}$ component as well as the size of the eight-sphere grow in the direction corresponding to the UV region of the dual quantum mechanics. This corresponds to the fact that the Hilbert space of the dual model grows indefinitely as we increase the energy. 

\subsection{Conformal quantum mechanics} %: AdS$_2\times S^2$ holography

If the theory allows for singular potentials, one can construct interacting theories with a scaling symmetry $t \to \lambda t$. Nevertheless, unitarity demands that deformations can be marginal at best. A simple example of this is the d'Alfaro-Fubini-Furlan (DFF) model \cite{deAlfaro:1976vlx}:
\begin{equation}
S = \int dt \left( \dot{x}^2  - \frac{g}{x^2} \right)~.
\end{equation}
The above model enjoys an $SL(2,\mathbb{R})$ conformal symmetry: 
\begin{equation}
t \to \frac{a t + b}{c t + d}~, \quad\quad x(t) \to {(c t + d)}^{-1} \, x \left( \frac{a t + b}{c t + d} \right)~.
\end{equation}
with $a$, $b$, $c$ and $d$ real and obeying $(ad-bc)=1$. The coupling $g$ here is exactly marginal and yields a unitary theory as long as $g> -\frac{1}{4}$. If the the coupling is further decreased the Hamiltonian ceases to be self-adjoint. This can be fixed through an appropriate self-adjoint extension at the price of a quantum mechanical RG flow \cite{Hammer:2005sa}. It is found that the coupling remains marginally relevant. As a consequence no further UV deformation can be achieved.
%Due to the conformal symmetry the high energy behavior of the model is intricately related to the low energy behavior. 

In the particular case of the DFF model, the analysis is complicated by the fact that the vacuum of the theory (once it is properly regulated in the infrared by the inclusion of a quadratic confining potential) breaks conformal invariance as it sits in a lowest weight representation of $SL(2,\mathbb{R})$ \cite{deAlfaro:1976vlx}. Recent developments in the study of SYK type models \cite{Sachdev:1992fk,pg,kitaev,Sachdev:2015efa,Anninos:2013nra,Anninos:2016szt,Polchinski:2016xgd,Maldacena:2016hyu,Maldacena:2016upp}, allow for the construction of theories where conformal invariance can be preserved in the vacuum to a better approximation.

In the situation where some large $N$ version of an $SL(2,\mathbb{R})$ invariant model has a geometric dual, one expects the UV part of the geometry to be different than that of the D0-brane geometry. 
As an example, we consider an AdS$_2\times S^2$ geometry with metric:
\begin{equation}\label{ads2throat}
ds^2 = \frac{-dt^2 + dz^2}{z^2} + d\Omega_2^2 = {-dt^2}{|\bold{x}|^2} + \frac{d\bold{x}^2}{|\bold{x}|^2}~. 
\end{equation} 
The dual theory is a quantum mechanics which (at least at large $N$) has an $SL(2,\mathbb{R}) \times SO(3)$ symmetry. The scaling symmetry is geometrized as $t \to \lambda t$ and $z \to \lambda z$. Unlike the D0-brane geometry, the emergent $S^2$ is $z$-independent. 
%This is consistent with scale invariance, nothing should depend on the scale which is roughly given by the $z$-coordindate. 
%Indeed, the boundary of the geometry is qualitatively different from the D0-brane case. 
We can consider more general asymptotically AdS$_2\times S^2$ configurations. For example, in Einstein-Maxwell theory the extremal throat can split into several ones carrying smaller charge \cite{Majumdar:1947eu,papapetrou}. For instance:
\begin{equation}\label{throats}
ds^2 = -\frac{dt^2}{\psi^2(\bold{x})} + \psi^2(\bold{x}) d\bold{x}^2~, \quad\quad \psi(\bold{x}) = \sum_{i=1}^n \frac{m_i}{|\bold{x}- \bold{x}_i|}~,
\end{equation}
is a solution to Einstein-Maxwell theory which approaches AdS$_2\times S^2$ for $|\bold{x}| \gg |\bold{x}_i|$ but splits into smaller AdS$_2\times S^2$ throats, each with their own horizon, whenever $\bold{x} \to \bold{x}_i$. Each of these horizons has a corresponding $S^2$ which is smaller than the asymptotic one. This is an indication of the reduced state space at lower energies. 
%Naturally, with $n=1$ and $\bold{x}_i = 0$ we obtain the usual geometry (\ref{ads2throat}).  The fragmented geometries break the $SL(2,\mathbb{R}) \times SO(3)$ symmetry. But a covariant structure remains. For instance $\bold{x} \to \lambda \, \bold{x}$, $\bold{x}_i \to \lambda \, \bold{x}_i$ and $t \to   t \, / \lambda$ remains a symmetry. So long as $\sum_i m_i^2$ is fixed to some constant, all these solutions are classically degenerate in energy. Whatever the quantum mechanics dual to AdS$_2\times S^2$ is, at large $N$ it should mimic the behavior of all these solutions. Small fluctuations include the infinitesimal classical motion of the tips, giving an effective $SL(2,\mathbb{R}) \times SO(3)$ invariant multi-particle mechanics of the degrees of freedom $\{\bold{x}_i(t) , \dot{\bold{x}}_i(t)\}$.
We see from $(\ref{throats})$ clearly that the size and shape of the asymptotic $S^2$ becomes $\bold{x}$ dependent, indicating that the full state corresponding to them is no longer conformally invariant, let alone $SO(3)$ invariant. 

In conclusion, in both models studied above we have noticed a very complicated IR structure present in quantum mechanical models. The UV, however, seems remarkably simple, parameterized by a small number of marginal deformations. This is in stark contrast to quantum field theory where one expects (naively) an infinite number of irrelevant deformations that can be described within effective field theory. 

%This means, there is an opportunity for us to exploit here. The strategy is to understand complicated IR dynamics from universal well understood UV quantum mechanical systems.

\subsection{Finite quantum mechanics?}

In the previous two examples, we have studied how two rather universal UV properties of quantum mechanics can manifest themselves geometrically. What about systems with finite dimensional Hilbert spaces? Such systems might be achieved, for example, by building the theory purely out of fermionic degrees of freedom \cite{Banks:2012ic,Verlinde:2016toy,Anninos:2015eji,Anninos:2016klf} or by introducing a Wheeler-de Witt like constraint on the spectrum by coupling the quantum mechanics to a worldline metric \cite{upcoming}. Perhaps a quantum mechanical system coupled to worldline gravity is of interest for describing a spacetime with no asymptotic boundary .

In this case, the energy spectrum caps both in the IR as well as the UV. Thus, to the extent that the energy is geometrized by some type of radial direction, we expect a geometry for which the size of space remains finite as a function of the holographic direction $r$ \cite{Anninos:2015eji}. One example of such a geometry is the static patch geometry of de Sitter space.\footnote{Another example is global AdS$_d$ with a large radial cutoff. In this case there is an additional parameter (the UV cutoff) unrelated to the number of colors, that tunes the number of states.} In this case, the cosmological horizon caps physics in the IR, and the shrinking celestial sphere caps the physics at the observer's worldline (at the South Pole of global de Sitter space) which we might identify as the UV. This can be seen from the metric itself:
\begin{equation}\label{sp}
\frac{ds^2}{\ell^2} = -dt^2 (1-r^2) + \frac{dr^2}{(1-r^2)} + r^2 d\Omega^2~,
\end{equation}
\noindent where $\ell$ is the dS curvature scale.
A basic challenge in interpreting the above metric holographically is that in the UV direction $r \to 0$, i.e. the direction for which time flows increasingly fast, the two-sphere caps to zero. (The capping of the sphere is often the behavior we observe in the deep IR region of a geometry dual to a gapped state.) Related to this, the $g_{tt}$ component of the metric does not grow parametrically, and hence it is hard to understand where the `UV operators' of the dual theory should be placed.\footnote{It seems worth pointing out a rather general relation in spherically symmetric sectors of general relativity for negative cosmological constant. The direction in which the proper size of the sphere grows seems to always correspond to the direction where the time-time component of the metric grows. This is compatible with the notion that deep inside the bulk one has integrated out degrees of freedom in the holographic theory. The relation between spheres and clocks remains valid so long as the matter content has positive energy. In fact, for the negative mass Schwarzschild black hole the time-time component of the metric grows in the direction where the sphere decreases, which is qualitatively similar to the behavior in the static patch.}
% (This issue is related to the fact that the cosmological horizon is observer dependent.) 
The issue is clear. Near $r=0$ there is no decoupling region where quantum gravity becomes non-dynamical. Therefore one might expect that coupling the dual quantum mechanics to gravity is crucial to have a UV complete description of the de Sitter static patch.

In what follows we will follow an alternative, novel and more concrete approach. We will describe the static patch by embedding  a dS$_2$ geometry into an asymptotically AdS$_2$ geometry. This will allow us to interpret the finite quantum mechanics dual to dS$_2$ as a subsector of the putative conformal quantum mechanics dual to AdS$_2$. This provides a natural UV completion that exploits the universal features described above. Due to the absence of a celestial sphere in the two dimensional case, dS$_2$ is slightly less confusing than its higher dimensional cousins. While we expect that this discussion can be generalized to that case, we will concentrate on this example in the remainder of this work. 

%Somehow the total number of states in the system becomes arbitrarily small\footnote{It is almost as if the states organize themselves into negative degrees of freedom inside the cosmological horizon and positive degrees of freedom outside, such that in the complete UV limit there are no degrees of freedom left at all. This might be related to the absence of a natural region where we can insert local operators from the quantum mechanics point of view.} as we increase the energy!

%In higher dimensions the full solution is dS$_2 \times S^2$ and the size of the sphere is $r$-independent. This is remarkably reminiscent of the (thermal patch of the) $SL(2,\mathbb{R})$ invariant AdS$_2\times S^2$ geometry. However in this case, there is no asymptotic boundary, instead there are two different IR repositories and which both connect to the same UV worldline at $r=0$. So it would seem to be a situation where we are gluing two different conformal quantum mechanics (possibly with a UV cutoff) and integrating over the boundary conditions. In Euclidean language we arrive at a similar picture. There we have that Euclidean dS$_2$ is a two-sphere, whereas Euclidean AdS$_2$ is the hyperbolic metric on the disk with some boundary conditions on the boundary $S^1$. Again, we can imagine gluing two disks together and integrating over the boundary conditions, two form a two-sphere. Notice, in accordance with our general considerations, that in de Sitter we have to integrate over the worldline metric also. 

\section{A/dS$_2$ and the centaur geometry}

In this section we consider dS$_2$ and AdS$_2$ geometries a little more carefully. We then discuss a geometry that looks like the static patch in the interior but has an AdS$_2$ boundary. We refer to this as the centaur geometry. 

\subsection{Static patch dS$_2$ geometry}

%\subsubsection{dS$_2$ and the worldline}

dS$_2$ is the simplest  geometry with a cosmological horizon. It is a solution of Einstein gravity with a positive cosmological constant, the Nariai geometry, which is dS$_2 \times S^n$ in $(n+2)$-dimensions. (Holography for the global Nariai geometry was investigated in \cite{Anninos:2009jt,Anninos:2009yc,Anninos:2010gh,Anninos:2011vd}.) The Nariai solution is quantum mechanically unstable. Indeed, if we allow for Hawking radiation one of the horizons eventually shrinks while the other grows. The end point of this process is the empty static patch geometry (\ref{sp}). One of the two horizons has negative specific heat, while the other has positive specific heat. In appendix \ref{nariaisec} we discuss the Nariai solution as a solution to a two-dimensional dilaton gravity theory.

The two-dimensional static patch geometry reads:
\begin{equation}\label{staticds2}
\frac{ds^2}{\ell^2} = -dt^2 (1-r^2) + \frac{dr^2}{(1-r^2)}~, \quad\quad\quad r \in (-1,1)~.
\end{equation}
Notice that there are two distinct horizons at $r = \pm 1$. The UV value of $r$, i.e. the one for which clocks tick the fastest, is at $r=0$. The geometry exhibits a discrete $\mathbb{Z}_2$ symmetry $r \to -r$.

If we go to Euclidean time $\tau = i t$, the dS$_2$ static patch becomes the round metric on the two-sphere. The two horizons now smoothly cap off at the poles of the sphere. Smoothness of the metric requires $\tau \to \tau + 2\pi$, indicating that the two horizons of the static patch are at a finite and equal temperature. The global extension of the coordinate system is:
\begin{equation}
ds^2 = -dT^2 + \cosh^2 T \, d\psi^2~,
\end{equation}
and has asymptotic boundaries in the infinite past and future where $T = \pm \infty$. The global geometry is $SL(2,\mathbb{R})$ invariant.  No single observer has causal access to the global geometry.

%Perhaps this is an indication that effectively negative degrees of freedom indeed play a role in constructing the static patch.

\subsection{Global AdS$_2$ geometry}

In global AdS$_2$, the metric is given by:
\begin{equation}
ds^2 = -dt^2 (1+r^2) + \frac{dr^2}{(1+r^2)}~, \quad\quad r \in \mathbb{R}~,
\end{equation}
which also exhibits an $SL(2,\mathbb{R})$ isometry group with generators:
\begin{eqnarray}
R &=& i\partial_t~,  \\
D &=&  -i\sqrt{1+r^2} \cos t \, \partial_r + i\frac{r}{\sqrt{1+r^2}} \sin t \, \partial_t~, \\
S &=&   -i\sqrt{1+r^2} \sin t \, \partial_r - i \frac{r}{\sqrt{1+r^2}} \cos t \, \partial_t~.
\end{eqnarray}
%\begin{equation}
%[D,R] = i S~, \quad [S,R] = -i  D~, \quad [S,D] = - i R~.
%\end{equation}
These satisfy the algebra $[S,R] = -i D$, $[D,R] = iS$ and $[S,D] = -i  R$. In addition there is a $\mathbb{Z}_2$ symmetry $r \to -r$. The generator of $t$-translations is the compact $U(1)$ subgroup of the $SL(2,\mathbb{R})$ isometry group, often denoted by $R = (H+a K)/2$. Here $K$ is the generator of special conformal transformations, $H$ is the Hamiltonian, $D$ is the dilatation generator and $S = (H-a K)/2$ the non-compact generator. The parameter $a$ is arbitrary but carries appropriate units for the expressions for $R$ and $S$ to make sense.

% that allows for a grading of the Hilbert space
Global AdS$_2$ has two asymptotic boundaries at $r = \pm \infty$ which are connected in the interior. This is related to the state operator correspondence in $(0+1)$-dimensions \cite{Strominger:1998yg,Sen:2011cn} where local operators of a theory on the Euclidean line $\mathbb{R}$ are mapped to the Hilbert space of a theory on $\mathbb{R} \times S^0$ where $S^0$ is now two points. The Lorentzian geometry is somewhat reminiscent of the eternal black hole in higher dimensions, however the two sides are now in causal contact from the point of view of the bulk. Thus, if there is any reasonable holographic interpretation of the global AdS$_2$ geometry, the CFTs seem to be interacting in a way that is stronger than simple entanglement of two disjoint Hilbert spaces. 
%
%Free scalar quantum fields for this geometry were studied in \cite{}. For fast falling boundary conditions near both boundaries, they exhibit $SL(2,\mathbb{R})$ boundary correlators of the form:
%\begin{equation}
%\langle \mathcal{O}(t) \mathcal{O}(t') \rangle = \left[ \frac{1}{\sin^2(t-t')} \right]^\Delta~.
%\end{equation}

If we go to Euclidean time $\tau = i t$, we have the Euclidean strip. Inclusion of the points at infinity gives us the Poincare disk. On the other hand, periodic identification of $\tau \sim \tau + \beta$ gives a smooth Euclidean geometry with two boundaries, which is locally equivalent to the hyperbolic disk. It differs globally from the disk, however, because the generator that we have identified is not compact for Euclidean AdS$_2$. The interpretation of this geometry is also somewhat mysterious, due to the disconnected boundaries. In the higher dimensional versions of AdS/CFT duality, the first state is identified with the AdS global black hole geometry while the second option corresponds to thermal AdS. The lack of horizon in the second case means that this geometry should correspond to a confined phase of the quantum mechanical theory and, therefore, with access to much fewer degrees of freedom than the black hole geometry.

Now consider an $SL(2,\mathbb{R})$ invariant quantum mechanics system, such as the DFF model (or some large $N$ vector or matrix generalization thereof \cite{Strominger:2003tm,verlinde}).  We could ask about the geometry dual to a single conformal quantum mechanics whose time evolution is governed by the $R$-Hamiltonian. When evolving with this Hamiltonian, the DFF model (and presumably most large $N$ generalizations thereof) has a unique normalizeable ground state that is not $SL(2,\mathbb{R})$ invariant.  The corresponding geometry dual to the ground state of a large $N$ DFF-like system will not have two asymptotic boundaries. Instead it will presumably end somehow in the deep interior. For an excited state with respect to this Hamiltonian, with sufficiently large entropy, the geometry might form a horizon. In such a case, the corresponding Euclidean geometry will have the topology of a disk rather than a cylinder. We now explore a particular example where such a situation occurs.  

%This is because the $R$-Hamiltonian introduces a scale.
%\subsection{Scalar waves}
%
%A set of normalizable wavefunctions is given by:
%\begin{equation}
%\phi(r,\omega) = \phi(\omega) \, P^\omega_{-1/2+\nu/2}(i r/L)~,
%\end{equation}
%with $\omega = (1/2+\nu)+n$ and $n \in \mathbb{Z}^+$. Euclidean global AdS$_2$ is the hyperbola with two disconnected circle boundaries. It is like coupling in the IR two Euclidean CQM, with Hamiltonian given by the compact generator $R = H+K$. In addition to an $SL(2,\mathbb{R})$ symmetry the above geometry has a discrete $\mathbb{Z}_2$ symmetry that allows for a grading of the Hilbert space. 
%
%

\subsection{Centaur geometry}

Consider a situation where the metric is given by:% for $r\ge 0$ is given by:
\begin{equation}
ds^2 = -\left(1+ \frac{r^3}{|r|}\right)dt^2  + \frac{dr^2}{\left(1+ \frac{r^3}{|r|} \right)}~, \quad\quad r\in(-1,\infty)~.
\end{equation}
%and for $r \in (-1,0)$ by:
%\begin{equation}
%ds^2 = -dt^2 (1-r^2) + \frac{dr^2}{(1-r^2)}~.
%\end{equation}
Such a centaur geometry is continuous and differentiable. The (possibly mixed) state corresponding to such a geometry in some putative dual quantum mechanics is not $SL(2,\mathbb{R})$ invariant. The left Hilbert space of excitations is that of (half) the two-dimensional static patch $\mathcal{H}_L$. The right Hilbert space is that of (half of) global AdS$_2$, $\mathcal{H}_R$. The Hilbert space of excitations is given by combining $\mathcal{H}_L$ and $\mathcal{H}_R$ in a smooth way. \begin{wrapfigure}{r}{5.5cm}
\includegraphics[scale=0.38]{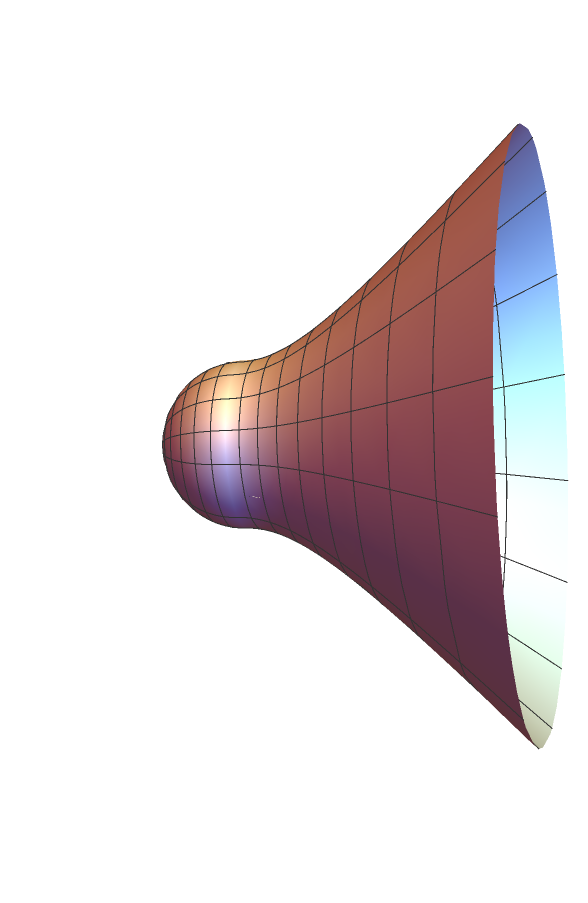}
\caption{\footnotesize{The Euclidean centaur geometry: a hemisphere merged to the hyperbolic cylinder.}}\label{centaurfig}
\end{wrapfigure} 
We can also consider the Euclidean centaur geometry, obtained by a Wick rotation of the Lorentzian one. Here, one is gluing a hemisphere to the two-dimensional hyperbolic geometry at the throat as shown in figure \ref{centaurfig}. We can view it as preparing a non $SL(2,\mathbb{R})$ invariant state.

%
%\begin{figure}
%\begin{center}\label{ratioplot}
%\includegraphics[scale=0.4]{centaurplot.pdf}
%\end{center}
%\caption{The Euclidean centaur geometry looks like a hemisphere merged to the hyperbolic cylinder.}
%\end{figure}

The centaur geometry is useful, in that it allows us to push the static patch worldline all the way to an AdS$_2$ boundary. In  this way, it allows us to avoid the issue of the observer dependence of the cosmological horizon since we have a preferred fixed worldline at the AdS$_2$ boundary. Moreover, it allows us to define a quantum mechanical holographic dual at the AdS$_2$ boundary where we can insert operators and compute wordline boundary correlation functions. The static patch is then interpreted as the IR physics of some excited thermal state in the QM dual. The entropy of the cosmological horizon is interpreted statistically as the entropy for that thermal state. 

In what follows we analyze the physical properties of centaur vacua in a simple dilaton two-dimensional gravity model. We could call the example above, where we are just gluing AdS$_2$ and dS$_2$, a sharp centaur. As we shall see, it is quite easy in dilaton gravity to find smooth centaur solutions as well.

%There is a jump in the vacuum energy at $r=0$.

%For a free massive field, the modes that are normalizeable at the boundary of AdS$_2$ are particular linear combinations of Legendre $P$ and $Q$ functions. Specifically:
%\begin{equation}
%\phi_R(r,\omega) = P^{ \omega}_{-1/2+\nu/2}(i r) + \frac{2i}{\pi} \, Q^{\omega}_{-1/2+\nu/2}(i r)~, \quad\quad r \in (0,\infty)~,
%\end{equation}
%behaves as $r^{-1/2-\nu/2}$ at large $r$ for all $\omega \in \mathbb{R}$. Modes in the dS$_2$ region $r \in (-1,0)$ are given by:
%\begin{equation}
%\phi_L(r,\omega) = \alpha \, P^{i \omega}_{-1/2+\tilde{\nu}/2}( r) + \beta \, Q^{i \omega}_{-1/2+\tilde{\nu}/2}( r)~,
%\end{equation} 
%where again $\omega \in \mathbb{R}$. Smooth modes require that $\phi_L(0,\omega) = \phi_R(0,\omega)$ and $\phi_L'(0,\omega) = \phi_R'(0,\omega)$. This fixes $\alpha$ and $\beta$. 
%

%\newpage
%
%BELOW IS INCORRECT
%
%Introducing a field $\Lambda$, we can rewrite $Z$ as (WRONG!):
%\begin{equation}
%Z = \int \prod_{m\in \mathbb{Z}} d\Lambda(m) \, e^{\sum_{m\in\mathbb{Z}} \Lambda(m) \Lambda(-m) \, G_{\Delta =1/4}(m) + \phi(m)\Lambda(-m) }~,
%\end{equation}
%where $G_{\Delta =1/4}(m)$ is the thermal $SL(2,\mathbb{R})$ invariant correlator with weight $\Delta = 1/4$:
%\begin{equation}
%G_{\Delta =1/4}(m) =  \frac{\Gamma(m/2+1/4)}{\Gamma(m/2+3/4)}~.
%\end{equation}
%We imagine that in the original theory, $\phi$ also sources a dynamical field with weight $\Delta = 3/4$. Maybe it is the shadow of $\Lambda$.

\section{Dilaton gravity with centaur vacuum}

We can construct the centaur solution in an effective two-dimensional dilaton-gravity model.
% obtained by reducing the cosmological Einstein action on a two-sphere. The four-dimensional metric takes the form
%\begin{equation}
%ds^2 = \frac{1}{\sqrt{\phi}} g_{\mu\nu} dx^\mu dx^\nu + 4\phi d\Omega^2~.
%\end{equation}
The two-dimensional Lorentzian theory is: %(CHANGED NOTATION $\phi \to -\phi$)
\begin{equation}\label{dilgrav}
S_L = \frac{\phi_0}{16\pi G} \int d^2x \sqrt{-g} R - \frac{1}{16\pi G} \int d^2x \sqrt{-g} \left(  \phi R + \, \ell^2 V(\phi) \right)~.% \frac{1}{2\sqrt{\phi}}  - \frac{3s}{2} \sqrt{\phi} \right)~. %  \partial_\mu \phi \partial^\mu \phi +
\end{equation}
%where the two-sphere area is $16\pi \phi$ and we have rescaled the four-dimensional cosmological constant for convenience. 
%Positive cosmological constant corresponds to $s=+1$ and negative to $s=-1$. 
The $\phi_0$ is a constant part for the full dilaton field $\Phi$, such that the full dilaton is given by $\Phi = \phi_0 - \phi$. In the second term, $\ell^2$ is a dimensionful constant playing the role of the cosmological constant. The first term in the action is purely topological due to the Gauss-Bonnet theorem in two dimensions. When equipped with the Gibbons-Hawking term, it computes the Euler characteristic. Since $\Phi$ is the difference between $\phi_0$ and $\phi$, the effective Newton constant of the non-topological piece has the opposite sign for the $\phi R$ piece of the action. All we require is $(\phi_0 - \phi) > 0$, but $\phi$ itself being a small deviation from $\phi_0$ can have either sign. We this choice of signs the geometry enters a strong coupling regime in the $\phi \rightarrow \infty$ region. This can be dealt with by using a UV regulator as we explain below. 

The equations of motion are given by:
\begin{eqnarray} \label{scalar}
8\pi G \, T^\phi_{\mu\nu} \equiv \nabla_\mu \nabla_\nu \phi - g_{\mu\nu} \nabla^2 \phi +\frac{\ell^2}{2} g_{\mu\nu}V(\phi) &=& 0~, \\ \label{ricci}
R +  \ell^2 \, V'(\phi) &=& 0~. 
\end{eqnarray}
%Notice that for $s = +1$ there is a solution with a constant $\phi = 1/3$. For this solution the two-dimensional metric is dS$_2$. There is no such solution for $s=-1$. The other solution has running dilaton. 
Due to diffeomorphism invariance, the stress tensor is conserved. Moreover, given the absence of a dynamical graviton in two-dimensions, the dilaton stress tensor vanishes identically. In fact, there are no propagating degrees of freedom in the dilaton gravity theory.

Using a B\"{a}cklund transformation \cite{Cavaglia:1998xj}, the general solution can be written as:
\begin{equation}\label{2dgeometry}
ds^2 = -N(\phi) dt^2 + \frac{d\phi^2}{\ell^2 N(\phi)}~,
\end{equation}
\noindent where:
\begin{equation}
N(\phi) = \int^\phi_{\phi_h} d\varphi \, V(\varphi) \, .
 \end{equation}

Here, we have used the running dilaton as a spatial coordinate in our geometry. $\phi_h$ parameterizes  an integration constant which will control the position of a horizon for this geometry. We can check that the Ricci scalar for (\ref{2dgeometry}) is given by $R = -\ell^2 \partial^2_\phi N(\phi)$. Clearly (\ref{ricci}) is solved by $N(\phi)$. As for (\ref{scalar}), using the relevant Christoffel coefficients one ends up with the equation $\partial_\phi N(\phi) = V(\phi)$, which is again satisfied by our solution. 

We will be interested in theories where $N(\phi)$ is a monotonic function, so the time lapse evolves in a particular direction determined by the RG flow in the $\phi$ direction. In this setup, $\phi_h$ represents the end of space in the Euclidean geometry. Depending on the behavior of the potential, it signals the presence of a horizon in the Lorentzian geometry. Our coordinates, therefore, span the interval $ \phi \in (\phi_h, \infty)$.

% = \left( \sqrt{\phi} - s \, \phi^{3/2} \right)$. For a more general potential, we can have a different situation. 
Our objective will be to construct smooth centaur geometries that interpolate between an AdS$_2$ boundary and a dS$_2$ cosmological horizon deep inside. To construct this geometry we would like a potential that is asymptotically linear in $\phi$ but with opposite slopes. This way we can attain constant Ricci scalars (with opposing signs) near the horizon and asymptotically. Let us define an infrared scale $\ell \epsilon$, where $\epsilon$ is a positive dimensionless constant, over which the potential interpolates between these two regimes. Then, we can get centaur geometries easily from any function $V(\phi)$ under the following simple conditions
\begin{equation}
V(\phi) \sim \left\{
     \begin{array}{cr}
    2 \phi  \quad\quad \textrm{if} \quad \phi \gg \epsilon\\
    -2 \phi  \quad\quad \textrm{if} \quad \phi \ll -\epsilon\\
   \end{array} \right. \quad\quad \textrm{and} \quad\quad \phi_h \ll -\epsilon
\end{equation}
\noindent 
where the factor of 2 has been chosen to match the standard notation, making $\ell^2 >0$ the curvature scale in the AdS$_2$ asymptotic region. As a concrete example, we could take
\be\label{example}
 V(\phi) = 2 \left(\sqrt{\phi^2+\epsilon^2}-\epsilon\right) .
 \ee
%
%\begin{equation}
%
%\end{equation}
%
%Then we have:
%\begin{equation}\label{nphi}
%N(\phi) =  \alpha^2 \left(\phi \sqrt{\phi^2+\epsilon ^2}-\phi_h \sqrt{\phi_h^2+\epsilon ^2}+\epsilon ^2 \log \frac{\phi + \sqrt{\phi^2+\epsilon^2}}{\phi_h + \sqrt{\phi_h^2+\epsilon^2}}-2 \, \left(\phi - \phi_h\right)  \, \epsilon \right)~,
%%\left[\phi \sqrt{\epsilon^2+\phi^2}+\epsilon^2 \log \left(\sqrt{\epsilon^2+\phi^2}+\phi\right)\right]~,
%\end{equation}
In other words, for $\phi_h < \phi \ll -\epsilon$, the two-dimensional metric looks to good approximation like dS$_2$ and like AdS$_2$ for  $\epsilon \ll \phi < \phi_b$. Here we assume $\phi_b$ is a UV cutoff where we can perform holographic renormalization as is the usual procedure in AdS/CFT computations. We assume throughout $\phi_0 \gg \phi_b$ so the the solution remains gravitationally weakly coupled in the whole geometry. In this particular example, the dilaton is running in a monotonically increasing fashion from the horizon to the near AdS$_2$ boundary. It is equal to the radial coordinate. Finally, the potential must have a minimum to interpolate between these two regimes. We pick the arbitrary value  $\phi = 0$ for this minimum.  Furthermore, we assume the potential is smooth and quadratic near the minimum,\footnote{This is not critical to the argument. Other analytic minima can be considered, but they require further fine tuning.} and we denote $V''(0) \equiv 2 \zeta^{-1}> 0$. This yields, in principle, a new infrared scale, $\ell \zeta$. In a simple case where there is only one infrared scale $\zeta \sim \epsilon$, as in (\ref{example}).
%While it can be useful to keep a concrete form in mind for the potential, as the one in (\ref{example}), we will be completely general and obtain universal expressions.  T
There is only one more property of the potential we will use for convenience: that the potential is strictly non-negative and that it vanishes quadratically at the single location $\phi=0$. %This guarantees the existence of zero temperature solutions. 

When the dS$_2$ geometry comes from the Nariai solution, one horizon is associated to the de Sitter horizon while the other to a black hole horizon which is degenerate in size to the cosmological one. As we discuss in appendix \ref{negdil}, whether the centaur is describing half of the dS$_2$ static patch containing the cosmological horizon, or the black hole depends on the behavior of the dilaton. The discussion in the main text corresponds to a dS$_2$ with a de Sitter like horizon. In appendix \ref{negdil} we briefly discuss the case where the potential is flipped and the boundary value of the dilaton is negative, where the dS$_2$ horizon becomes black hole like. Perhaps a way to obtain the full dS$_2$ involves merging the two types of centaur geometries, with black hole and de Sitter like horizons, across their AdS$_2$ boundary.
% We will investigate this in a forthcoming publication.
% It is also worth noting that we could have considered a different branch for which 

Finally, the parameter $\phi_h$ is physical here; one cannot simply rescale it away. In the case of pure (A)dS$_2$ the parameter $\phi_h$ would not have been physical. Because of the existence of this parameter we can dial the temperature of the horizon and study thermodynamic properties. 
 
%The dilaton is essentially constant up until $\phi \sim \phi_0$, which we take to be large. 
% (In recent parlance, the AdS$_2$ is in fact a nearly AdS$_2$.)

\subsection{Centaur thermodynamics}

The total Euclidean action is given by:
\begin{equation}
S_E = S_\phi + S_b+ S_0 ~.
\end{equation}
\noindent where
\be
S_\phi=\frac{1}{16\pi G} \int d^2x \sqrt{g} \left( \phi R +\ell^2 V(\phi) \right)
\ee
\noindent corresponds to the running dilaton action and 
\begin{equation}
S_b =   \frac{1}{8\pi G} \,  \int d\tau  \sqrt{h_{\tau\tau}}  \, \phi_b \, K~,
\end{equation}
\noindent is a  Gibbons-Hawking boundary term $S_b$, allowing for a well defined variational principle. Here $K$ is the trace of the extrinsic curvature at the boundary and the induced metric on a constant-$\phi$ surface is $h_{\tau\tau}(\phi) = g_{\tau\tau}(\phi)$. Lastly there is also the constant dilaton contribution to the action:
\begin{equation}
S_{0} =  - \frac{\phi_0}{16\pi G} \int d^2x \sqrt{g} R  -  \frac{\phi_0}{8\pi G} \,  \int d\tau  \sqrt{h_{\tau\tau}}  \, K~.
\end{equation}
The Euclidean solution is given by:
\begin{equation}\label{esol}
ds^2 =  N(\phi)  d\tau^2+ \frac{d\phi^2}{\ell^2 N(\phi)}~.
\end{equation}
The range of $\phi$ is  $\phi \in (\phi_h,\phi_b)$. We can expand $N(\phi)$ near the horizon as:
\begin{equation}
N(\phi) \sim V(\phi_h) ( \phi-\phi_h)
\ee
The Euclidean geometry requires the Euclidean time $\tau $ to be periodically identified with period:
\begin{equation}\label{tempphi}
\beta \equiv \frac{1}{T} = \frac{4\pi}{\ell V(\phi_h)} \sim  \frac{2\pi}{\ell |\phi_h|}~.%=  \frac{2\pi}{\alpha^2 \left(\sqrt{\phi_h^2+\epsilon^2}-\epsilon\right)} 
\end{equation}
\noindent where in the last equality we show the result in the regime $|\phi_h| \gg \epsilon$ (high temperatures). We see that $\phi_h$ can be interpreted as a (dimensionless) temperature of the system.
Thus, our solution has a finite temperature for $\phi_h$ non-zero. We can see in the expressions above that we have adjusted the zero of the potential to allow for zero temperature solutions $\beta \rightarrow \infty$ at $\phi_h \rightarrow 0$. Notice also that for a given temperature there exist two solutions corresponding to opposite sings for $\phi_h$. As long as $|\phi_h| \gg \epsilon$ these will look like dS$_2$ horizons or AdS$_2$ black holes, depending on the sign of $\phi_h$. We will show below that the dS$_2$ solution dominates the thermodynamics.

As long as the potential vanishes quadratically at some point (which we take to be at $\phi=0$) a zero temperature solution exists and the interior of the geometry is approximated by the following form:
\begin{equation}
ds^2 = \frac{\phi^3}{3 \zeta} d\tau^2 + \frac{3 \zeta}{\ell^2 \phi^3} {d\phi^2}~.
% = -\frac{\alpha^2 \phi^3}{3\epsilon}dt^2  + 3\epsilon \, \frac{d\phi^2}{\alpha^2 \phi^3}~.
\end{equation}
This corresponds to an asymptotically flat infinitely deep Euclidean cylinder in the IR. The solution is still asymptotically AdS$_2$ in the UV. One can think of this solution as a decoupling limit between the two sides of the global AdS$_2$ geometry where the bridge connecting them has become infinitely long. 

Using the Euclidean equations of motion, we can reduce the bulk part of the on-shell action to:
\begin{equation}
S_{\phi} =  \frac{\ell}{16\pi G} \, \int d \tau \int_{\phi_h}^{{\phi_b}} \, d \phi \left( - \phi \, V'(\phi) + V(\phi) \right)~.
\end{equation}
We also need the value of the extrinsic curvature at the AdS$_2$ boundary. The unit normal vector to a constant-$\phi$ surface is $n^\mu = \left(0,\sqrt{h_{\tau\tau}(\phi)}\right)$. We find:
\begin{equation}
K = \frac{1}{2} \, h^{\tau\tau}(\phi) \mathcal{L}_{n^\mu} h_{\tau\tau}(\phi) =  \frac{1}{2} \, \frac{\ell^2 V(\phi)}{\sqrt{h_{\tau\tau} (\phi)}}~.%\partial_\phi h_{\tau\tau}(\phi)~.% -\frac{1}{2\sqrt{g_{\tau\tau}}} \partial_\phi h_{\tau\tau}(\phi) = -\frac{1}{2 \sqrt{g_{\tau\tau}} }
\end{equation}
Putting everything together:
\begin{equation}\label{cZ} %- S_E=
\log Z[\beta] =  \frac{\phi_0}{4 G}- \frac{\beta \ell}{16 \pi G}\left[ 2 N(\phi_b) +\phi_h\, V(\phi_h) \right]~,
%-\beta \left[ \epsilon ^2 \log \left(\frac{\left(\sqrt{\beta ^2 \epsilon ^2-4 \pi ^2}+2 \pi \right) \left(\sqrt{\epsilon ^2+\phi_b^2}+\phi_b\right)}{\beta  \epsilon ^2}\right) +  \phi_b^2 \right]~.
% - \beta   \left[  \epsilon^2 \,\log\frac{\left(\sqrt{\epsilon ^2+ \phi_b^2}+{\phi_b}\right) \left(\sqrt{\epsilon ^2+ \phi_h^2}+{\phi_h}\right)}{\epsilon ^2} +  \phi_b^2 \right]~,
% = - S_{onshell} - S_b =  \left[ \epsilon^2 \,  \log \left( \frac{2  \phi_b \left( {c} + \sqrt{c^2+\epsilon^2} \right) }{\epsilon^2} \right) +  \phi_b^2 \right]~,
\end{equation}
Notice that the $N(\phi_b)$ term diverges. This is a familiar situation regarding UV divergences in QFT path integrals at finite temperature. The usual prescription \cite{Witten:1998zw} consists in substracting the zero temperature contribution, after properly adjusting the size of the thermal circle. This coincides with \cite{Grumiller:2007ju}, where it was proposed to add an additional boundary term to absorb this divergence\footnote{See also \cite{Gegenberg:1994pv,Grumiller:2006rc}.}. After regularizing we obtain:\footnote{Notice that we have not cancelled the contribution from the topological term when regularizing.}
\bea
 \log Z_{reg}[\beta]
% &=& -\lim_{\phi_b\rightarrow \infty} \frac{\beta \ell}{16 \pi G}\left( 2 N(\phi_b) +\phi_h\, V(\phi_h) - 2 \left(N(\phi_b)-N(0) \right) \left({\frac{N(\phi_b)}{N(\phi_b)-N(0)}}\right)^{1/2} \right) \nonumber\\
=  \frac{\phi_0-\phi_h}{4 G} -\frac{\beta\ell}{16 \pi G} \int^0_{\phi_h}  d\varphi \, V(\varphi) \, ,
%\frac{\phi_0}{4 G}- \frac{\beta}{16 \pi G}\left[ \phi_h\, V(\phi_h) + N(0)\right]=
\eea
\noindent where we have used the relation between $\phi_h$ and the temperature (\ref{tempphi}) and the form of $N$ in terms of the potential $V$.

Given the thermal partition function, we can compute thermodynamic quantities. For instance, the specific heat is:
\begin{equation}
C[\beta] = \beta^2 \partial^2_\beta \log Z_{reg}[\beta] = -\frac{1}{4 G} \frac{V(\phi_h)}{V'(\phi_h)}=\frac{\ell^2}{4 G} \frac{V(\phi_h)}{R(\phi_h)} ~.
%= \frac{8 \pi ^3}{\beta  \sqrt{4 \pi ^2-\beta ^2 \epsilon ^2}}~.
%\frac{8 \pi ^3 \beta  \epsilon^2 }{\left(4 \pi ^2 - \beta ^2 \epsilon^2 \right)^{3/2}} > 0~. %= \beta^2 \partial_\beta^2 \log Z[\beta]  
\end{equation}
Given that we are considering strictly positive potentials the specific heat of the solution is positive as long as the curvature at the horizon is positive. This singles out a negative value of $\phi_h$ for any potential with a single minimum at $\phi=0$. de Sitter like solutions (in the IR) are, therefore,  stable thermal saddles. At high temperatures $\beta \ell \ll 1/\epsilon$
\be
C[\beta] \sim \frac{1}{4 G} \frac{2\pi}{\beta \ell} \, ,
\ee
One can interpret this result as the appearance of more quantum mechanical degrees of freedom as one goes to higher energies. The result above predicts a constant density of states or equivalently the appearance of a linear mass spectrum of particles in the dual boundary theory. Meanwhile at low temperatures $\beta \ell \gg 1/\epsilon$, the heat capacity goes as
\be
 C[\beta] \sim \frac{1}{4 G} \sqrt{\frac{\pi \zeta}{\beta \ell}}\sim \frac{1}{4 G} \sqrt{\frac{\pi \epsilon}{\beta \ell}}~.% = \frac{1}{4 G} \sqrt{\frac{\epsilon \pi}{\beta \alpha^2}}  \, . 
\ee
\noindent where the expression after the second $\sim$ uses the assumption of just one infrared scale present in the problem.
This shows the presence of an even higher number of states at low temperatures. Notice that this is consistent with the curvature of spacetime becoming very low in the IR at low temperatures. In particular this matches the number of scattering states in a one dimensional potential, $g(E) \sim E^{-1/2}$, from the point of view of the boundary theory.\footnote{Notice that the above results are consistent with the quantum mechanics of a system of a large number of particles (``large $N$") given by different bound states of a pair of partons trapped in a flat potential in the UV that becomes quadratic at long distances. This result suggests important clues in terms of the operator content that should be present in a quantum mechanical dual to a de Sitter geometry.}

%This indicates a rich low energy density of states that goes as $1/\sqrt{E}$, which though singular is integrable.
%At $\beta = 2\pi/\epsilon$ it diverges, perhaps indicating the onset of a phase transition. This is in the regime of $c$ where the geometry no longer looks like the static patch anymore.
%Since this occurs at very small values of $\phi$, it may be sensitive to higher corrections of the potential 
%Notice that in a low temperature expansion, the leading pice of the specific heat is quadratic in the temperature.
%like that of the de Sitter horizon. 
The entropy receives a contribution from the topological piece which goes as $S = \phi_0/4G$. The correction to the entropy from the non-topological part of the action is given by:
\begin{equation}
\delta S[\beta] = \left(1- \beta \partial_\beta \right) \log Z_{reg}[\beta] = -\frac{\phi_h}{4G}~.
% = \frac{2 \pi  \sqrt{4 \pi ^2-\beta ^2 \epsilon ^2}}{\beta }
% -\frac{4 \pi \beta  \epsilon^2 }{\sqrt{{4 \pi ^2}-\beta^2 \epsilon^2 }}~.
\end{equation}
Notice that $\delta S$ increases with temperature and it is positive for IR positively curved centaur solutions. Finally, we can compute the energy of the system as:
\be
E[\beta] = \frac{\ell N(0)}{16 \pi G} = \frac{\ell}{16 \pi G}  \int^0_{\phi_h}  d\varphi \, V(\varphi) \,
\ee 
It is positive and it grows with the temperature.

In summary, we have found that solutions to the theory (\ref{dilgrav}) with infrared positive curved geometries, i.e. centaur solutions, are well behaved and stable thermodynamically. Curiously enough, it is negatively curved solutions in the IR that are unstable. In all cases the UV geometry is AdS$_2$ and the usual holographic dictionary allows calculation of QFT dual observables.

\section{Centaur waves}\label{quasiwaves}

In this section we compute the two point correlation functions in the boundary dual theory by following the AdS/CFT correspondence. We consider a bulk free scalar $\Psi$ of mass $\mu \ell$, where $\mu$ is a new dimensionless parameter. We will comment on general features of this problem and then we will perform this calculation exactly in the regime where we can disregard the interpolating region of the centaur geometry. That is, for a geometry which is constructed by gluing dS$_2$ to AdS$_2$ along the $\phi=0$ line: the sharp centaur. We show below that we can trust this calculation in the regime

\be
\epsilon \ell \ll 2 \pi T\, , \quad\quad \omega \ell \ll \frac{(2\pi T)^2}{\ell \epsilon}\, .
\ee

This is of particular importance as it allows a window into the region $2 \pi T \ll \omega \ell \ll \frac{(2\pi T)^2}{\ell \epsilon}$, where the de Sitter character of the geometry can be observed and separated clearly from the Rindler region. 

\subsection{Wave equation}

The wave equation for a free scalar of mass $\mu \ell$ is given by the Klein-Gordon equation in the centaur background. Expanding in Fourier modes $e^{-i\omega\ell t}$ we find:
\begin{equation}
\frac{\omega^2}{N(\phi)}  \Psi  + \partial_\phi \left( N(\phi) \partial_\phi  \Psi \right) = \mu^2 \, \Psi~,
\end{equation}
\noindent where $\omega$ is a dimensionless energy.

This problem is not easy to solve as there are many dimensionless parameters of relevance. It is a multi-scale problem. Let us do a quick recap of the dimensionless parameters entering the problem. These are energy scales expressed in terms of the AdS curvature scale. 
\begin{itemize}
\item $\mu$, the scalar field mass.
\item $\omega$, the energy of the excitation.
\item $\epsilon$, the infrared scale controlling the interpolating centaur region.
\item $|\phi_h|$, the Hawking temperature of the horizon.
\end{itemize}
The only relationship, so far, between these parameters is that we must have $|\phi_h| \gg \epsilon$ to guarantee our geometry is actually dS near the horizon.

As usual, when solving for a classical wave equation in gravity, the value of the effective gravitational constant does not play a role. Therefore there is no meaning in the normalization of $\phi$ and we can change coordinates to absorb one of the above scales. Under the change of coordinates $\hat \phi= \frac{\phi}{|\phi_h|}$ we obtain a rescaled wave equation:
\begin{equation}
\frac{\hat \omega^2}{N(\hat \phi)}  \Psi  + \partial_{\hat \phi} \left( N(\hat \phi) \partial_{\hat \phi}  \Psi \right) = \mu^2 \, \Psi~,
\end{equation}
\noindent where $\hat \omega = \frac{\omega}{|\phi_h|}$. Given that the only information we have about $V(\cdot)$ is that it grows linearly for large values of $|\phi|$, there is no need to rescale the potential after the change of coordinates, provided we now reidentify the infrared scale as $\hat \epsilon = \frac{\epsilon}{|\phi_h|}$. 
We are left with three dimensionless parameters:
\begin{itemize}
\item $\mu$, the scalar field mass.
\item $\hat \omega= \frac{\omega}{|\phi_h|}$, the energy of the excitation in units of the temperature.
\item $\hat \epsilon =\frac{\epsilon}{|\phi_h|} \ll 1$, the infrared scale in units of the temperature
\end{itemize}
In these coordinates the horizon is at $\hat \phi = -1$. Notice also that the condition $\hat \epsilon \ll 1$ implies that we know the precise form of the wave equation everywhere but in a small window in the $\hat \phi$ coordinate of order $(-\hat \epsilon, \hat \epsilon)$.

Now, in order to put the equation in Schr{\"o}dinger form,  it is convenient to introduce one last coordinate change:
\begin{equation}\label{ycoor}
y(\hat\phi) = \int^{\hat \phi}_0 \frac{ d \hat \varphi}{N(\hat \varphi)}~.
\end{equation}
The range of $y$ goes from $-\infty$ at the dS horizon to a positive constant $y_B =  \int^\infty_0 \frac{ d \hat \varphi}{N(\hat\varphi)} $ at the AdS boundary. In terms of $y$ the wave-equation simplifies to:
\begin{equation}\label{schro}
\left(-\frac{d^2}{dy^2} + \mu^2 \, N(\hat\phi(y))  \right) \Psi = \hat\omega^2 \, \Psi~.
\end{equation}
The potential is monotonically increasing with $y$ and tends exponentially to zero in the $y \to -\infty$ limit. 

Now let us introduce a scale $\hat \delta$ such that $\hat \epsilon \ll \hat \delta \ll 1$. This means we can solve the wave function exactly for $\hat \phi \in (-1,-\hat\delta) \cup (\hat \delta,\infty)$. As we take $\hat \delta \rightarrow 0$ all we need to do is match the wave-functions at this point. In order to determine the gluing conditions we integrate the Schr{\"o}dinger equation
\be\label{match}
-\left(\frac{\partial \Psi}{\partial y}(0^+)-\frac{\partial \Psi}{\partial y}(0^-)\right) = \hat\omega^2 \int_{-\delta}^\delta d\hat\varphi \frac{\Psi(\hat\varphi)}{N(\hat\varphi)}
\ee
Under our assumptions, $N(\hat\varphi)$ is a continuous finite function in this range as well as the wave function. That is why we tossed the contribution coming from the $\mu$ term in the equation. The term that remains could in principle contribute. This will only be the case if $\hat \omega^2 \gg 1$ to compensate for the small integral. We automatically reach the conclusion that for any finite $\hat \omega$ we can replace our problem for potential where the interpolating region has disappeared altogether. This is the approach we will follow momentarily.

It is instructive to see when this approximation fails. The term on the right can contribute for large $\hat \omega$. Here we can can go the WKB regime where the left hand side of equation (\ref{match}) is order $\hat \omega$. Therefore we can toss the right hand side if we can find $\hat \delta$ such that $\hat \epsilon \ll \hat \delta \ll 1$ and $\hat \omega\, \hat \delta \ll 1$. These conditions are guaranteed as long as:

\be\label{cond}
\hat \omega\, \hat \epsilon \ll1
\ee
For example, if $\hat \omega \,\hat \epsilon \sim \hat\epsilon^{\alpha} \ll 1$ for $\alpha>0$ then one can always find $\hat \delta \sim \hat \epsilon^{1-\gamma}$ where $0<\gamma<\text{min} (\alpha,1)$.

This is what we expected. For the gluing to be valid we need to be far from the region where we can solve the whole problem by a single interpolating WKB solution. But the condition for validity of the WKB regime is that the frequency of oscillation is large compared to any feature of the potential. This is exactly the opposite regime to (\ref{cond}). Therefore, if $\hat \omega \, \hat \epsilon \gg 1$ one can always resort to WKB. We will comment on our expectations in this regime after considering the sharp centaur solution.

Going back to our original variables, we have just shown that the form of the interpolating solutions is not important and the sharp gluing solution can be used in the regime of interest:
\be
\epsilon \ll |\phi_h| \sim \frac{2 \pi T}{\ell}\, , \quad\quad \omega \ll \frac{|\phi_h|^2}{\epsilon} \sim \frac{(2\pi T)^2}{\ell^2 \epsilon}\, .
\ee
Notice that whenever the temperature is large compared to the infrared scale $\epsilon \ell$ we have access, in this approximation, not only to hydrodynamic data but to a large range of quasinormal modes, which we typically expect at $\omega \ell \gtrsim T$.

\subsection{Exact gluing solution: the sharp centaur}

For any finite $\hat \omega$ we can just consider the sharp centaur solutions in the $\hat \epsilon \ll 1$ limit necessary for $dS_2$ to emerge. In this case we can perform the integral (\ref{ycoor}) explicitly. We obtain:
\be
y(\hat \phi) = \left\{ \begin{array}{cc}
\arctanh(\hat \phi) \quad & \quad \textrm{if} \quad -1<\hat\phi<0~, \\
\arctan(\hat \phi) \quad & \quad \textrm{if} \quad 0<\hat\phi<\infty~, \\
\end{array} \right.
\ee
which gives the potential:
\be
N(y) = \left\{ \begin{array}{cc}
\frac{1}{\cosh^2(y)} \quad & \quad \textrm{if} \quad -\infty<y<0~, \\
\frac{1}{\cos^2(y)}  \quad & \quad \textrm{if} \quad 0<y<\frac{\pi}{2}~. \\
\end{array} \right.
\ee
For both the $\hat{\phi}\in(-1,0)$ and $\hat{\phi}>0$ regions one can obtain exact solutions to the wave equation. They are given by the associated Legendre functions:
\begin{equation}\label{gluesolns}
\Psi(\hat{\phi}) =  \left\{ \begin{array}{cc} P^{i{\hat{\omega}}}_{\Delta-1}(-\hat{\phi}) &\quad  \textrm{if} \quad \hat{\phi}\in(-1,0)~, \\  
\gamma(\hat{\omega}) P^{{\hat{\omega}}}_{\tilde{\Delta}-1}(i \hat{\phi}) + \beta(\hat{\omega})  Q^{{\hat{\omega}}}_{\tilde{\Delta}-1}(i \hat{\phi} ) & \quad  \textrm{if} \quad \hat{\phi}>0~, 
\\
\end{array} \right.
\end{equation}
where it is convenient to define:
\begin{equation}
\Delta \equiv \frac{1}{2} + \frac{1}{2} \sqrt{1- 4\mu^2}~, \quad\quad \tilde{\Delta} \equiv \frac{1}{2} + \frac{1}{2} \sqrt{1+4\mu^2}~.
\end{equation}
To obtain (\ref{gluesolns}) we have imposed that the wave is purely ingoing at the de Sitter horizon. We require the modes to be smooth at $\hat{\phi} = 0$ up to the first derivative. This relates $\gamma({\hat{\omega}})$ to $\beta({\hat{\omega}})$. A simple but tedious calculation reveals the functions $\gamma(\hat{\omega})$ and $\beta(\hat{\omega})$, whose explicit form is given in appendix \ref{appendixsigma}. The Green function, according to the AdS/CFT dictionary, can easily be found by looking at the assymptotic behavior of these functions and combining $\beta(\hat \omega)$ and $\gamma(\hat\omega)$ in the appropriate way. We comment below on this quantity in Euclidean frequency space.

For now, a more transparent window into the physics correspond to the quasinormal modes. All we need to do is to impose that the modes are also fast falling near the AdS$_2$ boundary. To do this, we expand the modes at large $\hat{\phi}$ to find the coefficient of the non normalizable piece, which goes as $\hat{\phi}^{\tilde{\Delta}-1}$. Requiring that this coefficient vanishes imposes:
\begin{equation}
\frac{\Gamma \left(\frac{1}{2} (2-\Delta -i \hat{\omega})\right) \Gamma \left(\frac{1}{2} (1+\Delta -i \hat{\omega} )\right) \Gamma \left(\frac{\tilde{\Delta}-\hat{\omega} }{2}\right) \Gamma \left(\frac{\tilde{\Delta}+\hat{\omega} }{2}\right)}{\Gamma \left(\frac{1}{2} ( 1-\Delta-i \hat{\omega})\right) \Gamma \left(\frac{1}{2} (\Delta -i \hat{\omega} )\right) \Gamma \left(\frac{1}{2} (1+\tilde{\Delta}-\hat{\omega} )\right) \Gamma \left(\frac{1}{2} (1+\tilde{\Delta}+\hat{\omega} )\right)} = -1~.
\end{equation}
The solutions to the above equation yield the spectrum of quasi-normal modes of the system. A plot of the quasinormal modes of the system in the complex plane can be seen in figure \ref{qnm1}.
\begin{figure}
\begin{center}\label{qnm1}
\includegraphics[scale=0.4]{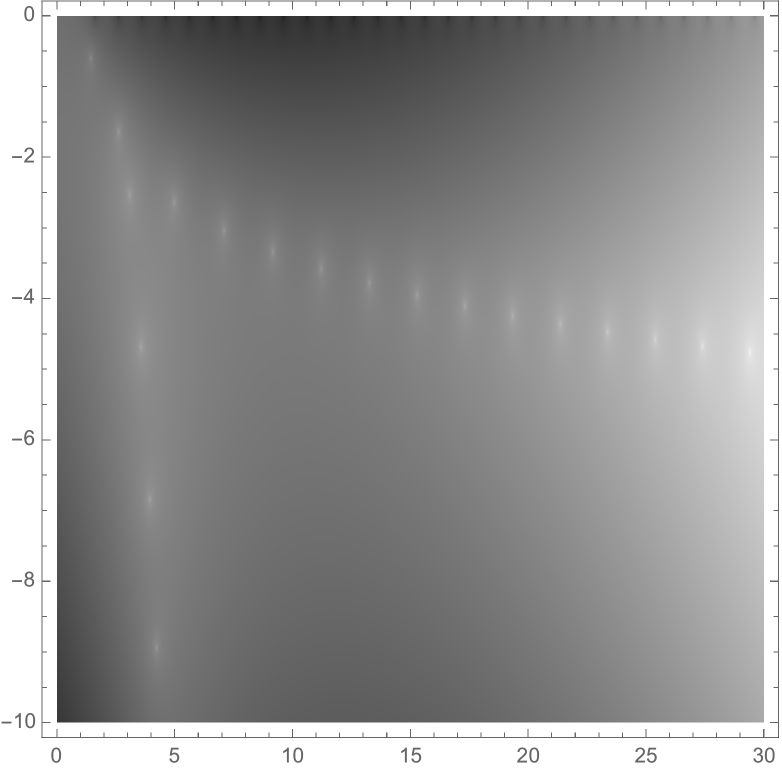}
\end{center}
\caption{Position of quasinormal modes in the complex $\hat \omega$-plane for $\mu=1$.}
\end{figure}
There are three regions of this plot that are worth highlighting:\footnote{We focus on the region with $\Re\left(\hat \omega\right) >0$ as the position of the poles is symmetric under $\Re\left(\hat \omega\right) \leftrightarrow -\Re\left(\hat \omega\right)$}

\begin{itemize}
\item \textbf{The long lived modes:} these are the modes where $\hat \omega \sim 1$. These are the longest lived modes in the system which correspond to energies (in the complex plane) of order $\omega \ell \sim 2 \pi T$. Notice that, generically, these modes have real and imaginary parts of comparable magnitude. Depending on the magnitude of $\mu$ they can be made more quasi-particle like ($\mu \gg 1$) or more dissipative ($\mu \ll 1$). These modes are localized around the interpolating region of the geometry.
\item \textbf{Quasi-particle modes:} these modes are associated to the AdS region of the geometry. Notice that they are not long lived, as their imaginary part is much larger than the temperature scale $T$. Nonetheless, their decay-rate is small compared to their energy. Concretely, for $n \gg 1 \in \mathbb{N}$,
\be
\omega \ell \sim 2 \pi \, T\left[ \tilde \Delta + 2 n - \frac{i}{\pi}   \log\frac{4 ( \tilde \Delta + 2 n)^4}{\mu^2} \right] \, .
\ee
The existence of quasi-particles is quite interesting as it suggests that some high energy excitations of the system must undergo a number of scatterings with the interpolating region before they can finally dissipate in the horizon. We trust the existence of these modes in the range (for the real part of the frequency) $2 \pi T \ll \omega \ell \ll \frac{(2\pi T)^2}{\ell \epsilon}$.
\item \textbf{Dissipative modes:} these modes are associated to the dS region of the geometry. They are dissipative as their imaginary part is much larger than their real part. In the regime (for the imaginary part of the frequency) $2 \pi T \ll \omega \ell \ll \frac{(2\pi T)^2}{\ell \epsilon}$ they are given by the expression, for $n \gg 1 \in \mathbb{N}$,

\be\label{centaurqnm}
\omega \ell \sim 2 \pi \, T\left[ -i \left(\frac{1}{2} + 2 n\right) + \frac{1}{\pi} \, \log\frac{ 8 \cosh\left(\pi \sqrt{\mu^2-\frac{1}{4}}\right)\left( \frac{1}{2} + 2 n\right)^4}{\mu^2}  \right]  \, ,
\ee
\noindent where we have quoted the result for masses $\mu^2 > \frac{1}{4}$ for simplicity.

Notice they have a real part as opposed to usual dissipative modes in the AdS$_2$ black hole. Somehow, in this geometry the deconfined phase is less efficient at thermalizing. Actually these modes have larger real parts than usual dS modes, which we discuss in the next section.

\end{itemize}

In conclusion, the physics in this case is radically different from finite temperature AdS geometries. In that case there is a confined phase with stable quasi-particle excitations, where decays are only given by loop effects. At high enough temperatures, AdS shows a deconfined phase where a horizon develops and all modes become totally dissipative. Centaurs are quite different. They appear to be deconfined phases (as a horizon is present), but dissipation is much less efficient. de Sitter like modes are still dissipative, and also show a real part to the frequency. Moreover, there are quasi-particle states. These long lived quasi-normal modes correspond to states having a hard time accessing the deep dS region in a way, perhaps, reminiscent of glassy physics.

\subsubsection{Expectations at $\omega \ell \gg {(2\pi T)^2}/{\ell \epsilon}$}

At high enough frequencies, our sharp centaur geometry is no longer a good approximation to any smooth centaur solution. In this limit, we expect a WKB function with trivial scattering matrix to solve the wave equation. The reason behind this, is that at high enough energies we can localize wave packets to distances much smaller than any feature of the geometry.
The upshot, is that in the AdS region we don't expect to find poles at high enough real frequency. Instead the series of poles found above should coalesce to form a cut in the Green function. Ultimately, this should reproduce the short distance behavior of the Poincare AdS Green function. On the other hand, in the dS region modes have only exponentially small support around the gluing region. Therefore we suspect that the usual spectrum of dS modes should persist in the mostly dissipative region.  

The conclusion is that centaur solutions, at high enough temperatures $2\pi T \gg \ell \epsilon$ show a dS behavior of quasinormal modes for imaginary frequencies $\omega \ell \gtrsim 2\pi T$, including crucially a real part that can only grow logarithmically. This provides a test of \textit{de Sitterness} for any quantum mechanical system. Notice the curious feature that, even though this is horizon physics, it is associated with the high temperature regime. The statement is that at high temperatures the system is somehow less dissipative than the usual deconfined phase associated to an AdS black hole.

\subsection{Boundary correlator}

It is also of interest to compute the boundary correlator of the free scalar in a fixed centaur background. For this it is convenient to go to Euclidean time $\tau \sim \tau + \beta$, for which the thermal frequencies become $\omega \ell = 2\pi T m$ with $m \in \mathbb{Z}$. The Euclidean action for $\Psi$ is given by:
\begin{equation}
S_E = \frac{1}{2} \, \int d\tau \int_{-1}^{\hat{\phi}_b} d\hat{\phi} \left( \partial_\nu \Psi \partial^\nu \Psi + \mu^2 \Psi^2 \right)~.
\end{equation}
On-shell, the action reduces to a boundary term, and we are interested in its value at large $\hat{\phi}_b$. Expanding in thermal frequency modes:
\begin{equation}
S_E =  \frac{1}{2} \, \lim_{\hat{\phi}_b \to \infty} \sum_{m \in \mathbb{Z}} (1+ \hat{\phi}_b^2)  \Psi(\hat{\phi}_b,m) \partial_{\hat{\phi}} \Psi(\hat{\phi}_b,-m)~.
%S_E =  \frac{1}{2} \, \lim_{\phi_b \to \infty} \sum_{m \in \mathbb{Z}} \alpha^2(\phi_h^2+\phi_b^2)  \Psi(\phi_b,m) \partial_\phi \Psi(\phi_b,m)~. %\approx \sum_{m \in \mathbb{Z}} 
\end{equation}
At large $\hat{\phi}$, the field behaves as:
\be
\Psi \approx \Psi_0(m) \left( \frac{\hat{\phi}^{\tilde\Delta-1} + \sigma(m)  \hat{\phi}^{-\tilde\Delta}}{\hat{\phi}_b^{\tilde \Delta-1} + \sigma(m) \hat{\phi}_b^{-\tilde \Delta}} \right)~,
\ee
where we have normalized it such that $\Psi(\hat{\phi}_b,m) = \Psi_0(m)$ is the boundary source for the dual operator $\mathcal{O}_\Psi$. The function $\sigma(m)$, whose exact form in terms of $\gamma(m)$ and $\beta(m)$ is given in appendix \ref{appendixsigma}, computes the non-local piece of the boundary two-point function of $\mathcal{O}_\Psi$, as can be seen from the on-shell action at large $\hat{\phi}_b$:
%$ Putting it all together:% \red{CHECK NORMALIZATION}:
\begin{equation}
S_E =  \frac{1}{2} \,  \sum_{m \in \mathbb{Z}} \, \Psi_0(m) \Psi_0(-m) \left( (\tilde\Delta -1) \, \hat{\phi}_b  -   \frac{\left( 2\tilde\Delta-1\right)}{\hat{\phi}_b^{2(\tilde\Delta-1)}} \, \sigma(m) + \ldots \right)~. %(1+r_c^2)  \phi(r_c,m) \partial_r \phi(r_c,m)~.
\end{equation}
%From this we can evaluate the boundary-to-boundary two-point function in the centaur state. 
Its exact form is not very illuminating, but we can easily plot it as a function of the thermal frequency. As a simple check, however, we can consider the massless case (which in two-dimensions in conformally coupled), where we find the boundary correlator $\sigma(m) \propto m$. More generally, we find numerically that $\sigma(m) \sim m^{2\tilde{\Delta}-1}$ {\it only} at large $m$. We show an example in figure \ref{ratioplot}. Thus, the two-point function is not scale invariant unless we go into the UV regime $m \gg 1$. This is consistent with the centaur geometry not being dual to an $SL(2,\mathbb{R})$ invariant state in the putative dual quantum mechanical model.
\begin{figure}\label{ratioplot}%{r}{5.5cm}
\begin{center}
{\includegraphics[scale=0.7]{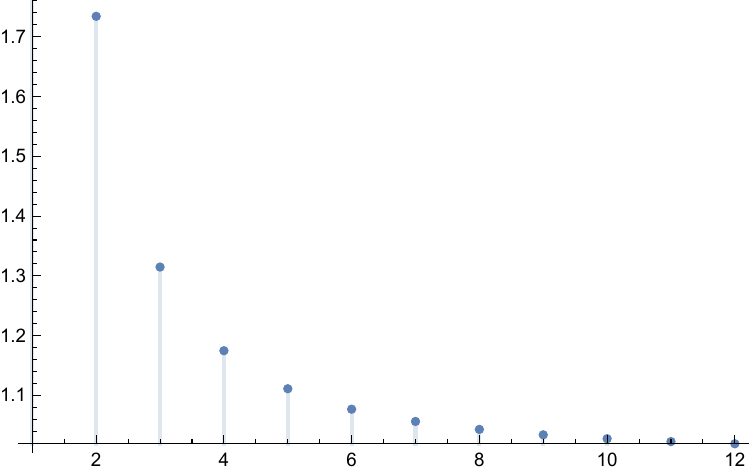}}
\caption{Plot of $\sigma(m)/(c \, m^{2\tilde{\Delta}-1})$ vs. $m$ for $\tilde{\Delta} = (1+\sqrt{17})/2 \approx 2.56$. The constant $c$ is chosen such that the curve flattens to one at large $m$.}
\end{center}
\end{figure}

%\begin{wrapfigure}{r}{5.5cm}
%\boxed{\includegraphics[scale=0.42]{ratio.pdf}}
%\caption{Plot of $\sigma(m)/m^{2\tilde{\Delta}-1}$ vs. $m$ for $\tilde{\Delta} = (1+\sqrt{17})/2 \approx 2.56$.}\label{ratioplot}
%\end{wrapfigure} %As $m$ increases, $\sigma(m)$ approaches the scale invariant two-point function.

%In figure \ref{qnmfig} we display
%\begin{figure}
%\begin{center}\label{qnmfig}
%\includegraphics[scale=0.5]{qnm1.jpg} , \quad \includegraphics[scale=0.5]{qnm2.jpg}
%\end{center}
%\caption{Density plot of retarded Green function in the complex $\omega$ plane for $\tilde{\Delta} = 1/2+\sqrt{37}/2 \approx 3.54$ and $\Delta = 1/2+ \sqrt{35} i /2 \approx 0.5+ 2.96 i $. The white region peaks at the value of the lowest quasinormal mode which is found numerically to be (left) $\omega_{qnm} \approx  3.44 - 0.498 i$. Another qusinormal mode (right) lies at $\omega_{qnm} \approx - 3.44 - 0.498 i$.}
%\end{figure}

% as:
%\begin{equation}
%\sigma(m) = ~.
%\end{equation}
%When pushed to the boundary the mode will include both fast and slow falling modes, i.e. there will be a source turned on. The partition function of the theory with such a source will be the same as that of the hemisphere when pushed all the way to $r = 0$. 

\section{A bestiary of worldline correlators}\label{ds2}

In this section we consider the correlations along the worldline in different settings and compare with the results of the previous section.

\subsection{dS$_2$ worldline correlators}

We consider the correlations along the worldline for a free scalar field in the fixed dS$_2$ static patch geometry (\ref{staticds2}). 
%In the Fourier representation, 
%\begin{equation}
%\Psi(r,t,\Omega) =  \int_{\mathbb{R}} d\omega  \left( \Psi_{\omega} (r) e^{-i\omega t}  + h.c. \right)~.
%\end{equation}
%we obtain the wave-equation:
% and we work in units where $\ell=1$
%
%\begin{equation}\label{eom2}
%\partial_r (1-r^2) \partial_r \, \Psi_\omega(r) + \frac{\omega^2}{(1-r^2)} \Psi_\omega(r) - \mu^2 \, \Psi_\omega(r) = 0~.
%\end{equation}
%The general solution is given in terms of associated Legendre polynomials:
%\begin{equation}
%\Psi_\omega(r) = \alpha \,  P_{\Delta-1}^{i\omega}(r) + \beta \, Q_{\Delta-1}^{i\omega}(r)~, \quad\quad \Delta \equiv  \frac{1}{2}+{\sqrt{\frac{1}{4}- \mu^2}}~.
%\end{equation}
%The wave-equation (\ref{eom2}) can be written more geometrically in terms of the $SL(2,\mathbb{R})$ isometries of dS$_2$. These are given by (define $r = \cos\theta$):
%\begin{equation}
%L_\pm =  e^{\pm t} \left[ \pm \partial_\theta - \cot \theta \,  \partial_t \right]~, \quad\quad L_0 = i \partial_t~.
%\end{equation}
%Thus, using the quadratic Casimir, we can write the wave-equation as:
%\begin{equation}
%\left[ \frac{1}{2} \left( L_+ L_- + L_- L_+ \right) + L_0^2 \right] \Psi = \Delta(1-\Delta)  \Psi~.
%\end{equation}
%where $\Delta$ can take either value below:
%\begin{equation}
%\Delta_\pm = \frac{1}{2} \pm {\sqrt{\frac{1}{4}-\mu^2}}~.
%\end{equation}
%We can generate each tower of quasinormal modes by acting on the $n=0$ mode with the raising operator $L_+$. 
%Thus each tower furnishes a highest weight representation of $SL(2,\mathbb{R})$.
It is convenient to introduce $y = \tanh^{-1} r/\ell$, with $y \in \mathbb{R}$ such that the wave-equation takes the Schr{\"o}dinger form:
\begin{equation}\label{eomy}
\left(-\frac{d^2}{dy^2} + \frac{\Delta(1-\Delta)}{\cosh^2 y}  \right) \Psi(y) = \hat{\omega}^2 \Psi(y)~. %\quad \Delta \equiv \frac{1}{2} + \frac{1}{2} \sqrt{{1}-4\mu^2}~.
\end{equation}
This is of the P{\" o}schl-Teller type. 
The solutions are conveniently organized in terms of modes which are purely oscillating in the positive and negative $y$-directions:\footnote{Our definition of the associated Legendre $Q$ polynomial is:
\begin{equation}\nonumber
Q_{\Delta-1}^{i\hat{\omega}}(\tanh y) = \frac{\left( \text{sech} y  \right)^{-i \hat{\omega} }  \, _2F_1\left[\frac{1}{2} (\Delta -i \hat{\omega}),\frac{1}{2} (\Delta+1 -i \hat{\omega});\Delta +\frac{1}{2};\frac{1}{\tanh ^2(y)}\right]}{(\tanh y)^{\Delta -i \hat{\omega}}}~,
\end{equation}
for $y<0$ and $\omega \in \mathbb{R}$. As can be checked explicitly, this is indeed a solution to the differential equation (\ref{eomy}).
}
\begin{eqnarray}
\Psi_{+}(y) &=& P_{\Delta-1}^{i\hat{\omega}}(\tanh y)~, \\
\Psi_{-}(y) &=& Q_{\Delta-1}^{i\hat{\omega}}(\tanh y) - \alpha(\hat{\omega}) P_{\Delta-1}^{i\hat{\omega}}(\tanh y)~, 
\end{eqnarray}
where $\alpha$ is given by:
\begin{equation}
\alpha(\hat{\omega}) = -\frac{2^{{\Delta-1} } e^{-2 \pi  \hat{\omega} -i \pi  (\Delta-1) } \Gamma \left(\Delta +\frac{1}{2}\right) \Gamma (1-\Delta -i \hat{\omega} )}{\sqrt{\pi }}~.
% -\frac{2^{\nu } (-1)^{-\nu +3 i \omega +1} \Gamma \left(\nu +\frac{3}{2}\right) \Gamma (-\nu -i \omega )}{\sqrt{\pi }}~.
%-\frac{\sqrt{\pi } 2^{\nu } e^{-2 \pi  \omega -i \pi  \nu } \Gamma \left(\nu +\frac{3}{2}\right) \Gamma (-i \omega ) \Gamma (-\nu -i \omega )}{\pi  \Gamma (-i \omega )-\sin (\pi  \nu ) \Gamma (i \omega ) \Gamma (-\nu -i \omega ) \Gamma (\nu -i \omega +1)}~.
%\alpha^{-1} = \text{sech}(\pi  \omega ) \left(-\frac{2 \, \sin (\pi  \nu )}{\pi }+\frac{2 \, \Gamma (-i \omega )}{\Gamma (i \omega ) \Gamma (-\nu -i \omega ) \Gamma (\nu -i \omega +1)}\right)~.
\end{equation}
At large positive $y$ we have $ \Psi_{+}(y) \sim e^{i\hat{\omega} y}$ whereas at large negative $y$ we have $\Psi_{-}(y) \sim e^{-i\hat{\omega} y}$. As a simple check, our solutions at $\Delta = 1$ (which is the conformally coupled case) reproduce the purely left- and right-moving flat space expressions.

Thus, we can compute the retarded Green's function defined by the boundary conditions that there be no incoming flux from the past horizons. We have:
\begin{equation}
G_R(y,y';\hat{\omega}) = \left\{ \begin{tabular}{c}
 $\frac{ \Psi_{+}(y) \Psi_{-}(y')}{w(\hat{\omega})}~, \quad\quad y > y',$ \\
 $\frac{ \Psi_{-}(y) \Psi_{+}(y')}{w(\hat{\omega})}~, \quad\quad y < y',$
\end{tabular}\
\right.
\end{equation}
where the Wronskian $w(\hat{\omega}) \equiv \Psi_+ \Psi_-' - \Psi_- \Psi_+'~$ is given by:
\begin{equation}
w(\hat{\omega}) = \frac{2^{\Delta} e^{-\pi  (\hat{\omega} +2 i (\Delta-1) )}}{\sqrt{\pi }} \, \frac{\Gamma \left(\Delta +\frac{1}{2}\right)}{ \Gamma (\Delta -i \hat{\omega} )}~,
%\frac{i \left(-\frac{1}{2}\right)^{-\nu -1} e^{-\pi  \omega } \,  \Gamma \left(\nu +\frac{3}{2}\right) \text{csch}(\pi  \omega ) \sin (\pi  (\nu +i \omega ))}{\sqrt{\pi } \, \Gamma (\nu -i \, \omega +1)}~.
%-\frac{i 2^{\nu +1} (-1)^{-\nu +i \omega } \Gamma \left(\nu +\frac{3}{2}\right) \text{csch}(\pi  \omega ) \sin (\pi  (\nu +i \omega ))}{\sqrt{\pi } \Gamma (\nu -i \omega +1)}~.
% \frac{i \, 2^{\nu +1} e^{-\pi  \omega -i \pi  \nu } \Gamma \left(\nu +\frac{3}{2}\right) \text{csch}(\pi  \omega ) \Gamma (i \omega ) \left(\frac{\pi }{\Gamma (i \omega ) \Gamma (-\nu -i \omega ) \Gamma (\nu -i \omega +1)}-\frac{\sin (\pi  \nu )}{\Gamma (-i \omega )}\right)}{\sqrt{\pi } \Gamma (\nu +i \omega +1)}~.
%-\frac{\sin (\pi  (\nu -i \omega )) \csc (\pi  (\nu +i \omega )) \Gamma (i \omega -\nu )}{\Gamma (-\nu -i \omega )}~.
\end{equation}
and is manifestly independent of $y$. 

At this point we can study the analytic structure of $G(y,y';\hat{\omega})$ in the complex $\hat{\omega}$-plane \cite{LopezOrtega:2006my,Cardoso:2003sw}. If we expand the Green function at $y \lesssim y' = 0$ we find the wordline correlator:
\begin{equation}\label{Gds2final}
G_R(\hat{\omega}) =  \frac{1}{4} \frac{\Gamma \left(\Delta_+/2 -i\hat{\omega}/2 \right)}{\Gamma \left(1-\Delta_+/2-i\hat{\omega}/2 \right)} \frac{\Gamma \left(\Delta_-/2 -i\hat{\omega}/2   \right)}{ \Gamma \left(1-\Delta_-/2 - i \hat{\omega}/2 \right)}~,
% \frac{\left(1+2 e^{i \pi  (\Delta_+ + i \omega )}\right)}{4} \frac{\Gamma \left(\Delta_+/2 -i\omega/2 \right)}{\Gamma \left(1-\Delta_+/2-i\omega/2 \right)} \frac{\Gamma \left(\Delta_-/2 -i\omega/2   \right)}{ \Gamma \left(1-\Delta_-/2 - i \omega/2 \right)}~.
\end{equation}
where:
\begin{equation}\label{eqds2}
\Delta_\pm = \frac{1}{2} \pm  \frac{1}{2} {\sqrt{{1}-4\mu^2}}~.
\end{equation}
The pole structure arises from the Gamma functions. Recall that $\Gamma(n)$ has a simple pole whenever $n$ is a non-positive integer. Thus we find that $G(\hat{\omega})$ has two towers of poles at:
\begin{eqnarray}\label{qnm}
\hat{\omega}^{(+)}_n &=& -i (2n+\Delta_+)~, \quad\quad n = 0,1,2,\ldots \\ \label{qnm2}
\hat{\omega}^{(-)}_n &=& -i (2n+\Delta_-)~, \quad\quad n = 0,1,2,\ldots
\end{eqnarray}
%Notice that the lowest lying quasinormal modes are given by $\omega_0 = -i/2 \pm \sqrt{\mu^2-1/4}$. 
At large $\mu$ and $n\gg 1$ these agree with those computed in (\ref{centaurqnm}). This is encouraging. It indicates that such boundary observables capture the fact that the interior of the centaur geometry is de Sitter like. 
%which are very near the ones we observed for a scalar of the same mass in the centaur geometry. 

%In appendix \ref{hemi}  we reproduce (\ref{Gds2final}) doing a calculation on the hemisphere.
%\footnote{Perhaps the way to understand the doubled $SL(2,\mathbb{R})$ structure of the pure static patch is as a residual structure starting from a the Nariai (unstable) fixed point flowing to the empty static patch.}

%As in the case of the higher dimensional static patch \cite{Anninos:2011af}, we find that the worldline correlator in frequency space is the product of two $SL(2,\mathbb{R})$ invariant correlators, one with conformal weight $\Delta_+/2$ the other with weight $\Delta_-/2$. We can take the Fourier transform of $G_0(\omega)$ and express it as a function of the worldline time coordinate. For the case $\mu_{eff}^2 = 1/4$ we have: 
%\begin{equation}
%G_0(t) = \int_\mathbb{R} d T \,  \left( \frac{1}{\sinh |T|} \right)^{1/2} \, \left( \frac{1}{\sinh |T+t|} \right)^{1/2}~.
%\end{equation}
%The short time singularity in the small $t$ limit goes as $\sim \log |t|$, as follows from the above expression. On the other hand, for large time separations $t \gg 1$ (recall we are working in units $\ell = 1$), we have $G_0(t) \sim e^{-t/2}$, or more generally $G_0(t) \sim e^{-\Delta_\pm t}$. 

\subsection{dS$_{d+1}$ worldline correlators}

For dS$_{d+1}$, for $d\ge 3$, the retarded Green function was computed in \cite{Anninos:2011af}:
\begin{equation}\label{GRd}
G_R(\hat{\omega}) =  \frac{\Gamma \left(\Delta_l -i(\hat{\omega}+\hat{\omega}_Q)/2 \right)}{\Gamma \left(1-\Delta_l-i(\hat{\omega}+\hat{\omega}_Q)/2 \right)} \frac{\Gamma \left(\Delta_l-i(\hat{\omega}-\hat{\omega}_Q)/2   \right)}{ \Gamma \left(1-\Delta_l - i(\hat{\omega}-\hat{\omega}_Q)/2 \right)}~,
\end{equation}
where now:
\begin{equation}
\Delta_l = \frac{d}{4} + \frac{l}{2}~, \quad\quad \hat{\omega}_Q =  \frac{1}{2} \sqrt{4\mu^2 - {d^2}}~.
\end{equation}
Note that $G_R(\hat{\omega})$ also depends on the angular momentum $l \in \mathbb{N}$. Remarkably, the form of $G_R(\hat{\omega})$ is essentially the same as that for the two-dimensional case, regardless of the dimensionality of space. The main difference is that there is now a tower of weights labelled by $l$. The quasinormal modes are given by:
\begin{equation}\label{omqnm}
\hat{\omega}^{(\pm)}_{n} = - 2 i \left( \Delta_l + n  \right) \pm \hat{\omega}_Q~, \quad\quad n =0,1,2,\ldots
\end{equation}
Notice that increasing $l$ increases the imaginary part of $\hat{\omega}^{(\pm)}_n$ and hence higher angular momentum modes are increasingly dissipative. 

The structure of (\ref{omqnm}) is in stark contrast to quasinormal modes in asymptotically flat or AdS space. In dS, large $l$ quasinormal modes are dissipative and fall into the cosmological horizon. A very heavy particle\footnote{But not too heavy that it becomes a black hole filling the whole static patch!} in dS (with $\mu^2 \gg 1$) contributes a large real part to the quasinormal modes, since it safely sits on the worldline far away from the cosmic horizon. For AdS or flat space black holes, $l$ contributes to the real part of the quasinormal mode, indicating that the states become less dissipative. Indeed, in AdS or flat space, a quasinormal mode with large $l$ explores the asymptotic boundary of space. A heavy particle falls into the black hole and hence is highly dissipative, i.e. the mass contributes to the imaginary part in AdS/flat space. However, for AdS$_5\times S^5$ the angular momentum modes of the $S^5$ contribute to the imaginary part of the quasinormal modes (similar to the $l$ contribution in dS), since they appear as massive particles in the AdS$_5$. The $S^5$ is emergent, and not part of the field theory directions of the dual CFT. 
%We take the above observations as a hint that the higher dimensional story will not be too different from the dS$_2$ case analyzed in this paper. 
%Again this is in contrast to the asymptotically flat or AdS black hole.

\subsection{The SYK spin-fluid state}

Here we note that (\ref{Gds2final}) is given by the product of two SYK propagators.
%As we have seen, for large enough masses, the dS$_2$ wordline correlator is given by:
%\begin{equation}\label{Gds2}
%G_0(\omega) %&=&  \frac{1}{4} \frac{\Gamma \left(\Delta_+/2 -i\omega/2 \right)}{\Gamma \left(1-\Delta_+/2-i\omega/2 \right)} \frac{\Gamma \left(\Delta_-/2 -i\omega/2   \right)}{ \Gamma \left(1-\Delta_-/2 - i \omega/2 \right)}~\\
%=  \frac{1}{4} \, \frac{\Gamma \left(\Delta -\frac{i(\omega-\omega_Q)}{2\pi T} \right)}{\Gamma \left(1-\Delta - \frac{i(\omega-\omega_Q)}{2\pi T}  \right)} \, \frac{\Gamma \left(\Delta -\frac{i(\omega+\omega_Q)}{2 \pi T} \right)}{\Gamma \left(1-\Delta - \frac{i(\omega+\omega_Q)}{2 \pi T}  \right)}~,
%%G_0(\omega) = \frac{\left(1+2 e^{i \pi  (\delta + i \omega )}\right)}{4} \frac{\Gamma \left(\Delta_+/2 -i\omega/2 \right)}{\Gamma \left(1-\Delta_+/2-i\omega/2 \right)} \frac{\Gamma \left(\Delta_-/2 -i\omega/2   \right)}{ \Gamma \left(1-\Delta_-/2 - i \omega/2 \right)}~.
%\end{equation}
%with $\omega_Q$ real. Explicitly,
%\begin{equation}
%\omega_Q = \ \sqrt{\mu^2-\frac{1}{4}}~, \quad\quad T= \frac{1}{\pi}~,\quad\quad \Delta=\frac{1}{4}~.%\Delta_\pm = \frac{1}{2} \pm \frac{\sqrt{1-4m^2_{eff}}}{2}~, \quad\quad 
%\end{equation}
%At this point it is useful to recall the retarded Green function of an $SL(2,\mathbb{R})$ invariant system finite temperature $T$. For an operator of weight $\Delta$, this is given by (setting $\hbar = k =1$):
%As mentioned in the introduction, several works have recently considered a class of large $N$ models with Green's functions of the form (\ref{Gds2}). 
The particular model of interest \cite{Sachdev:2015efa} consists of $N$ randomly interacting complex fermions charged under a global $U(1)$. The specific interaction for $\Delta = 1/4$ is given by a purely quartic interaction. These models, known as the SYK models, exhibit an emergent scaling symmetry at low temperatures and the low frequency retarded Green's function is computed as:
\begin{equation}\label{Gret}
G_R(\hat{\omega}) = \frac{\mathcal{N} e^{-i\theta}}{(2\pi T)^{1-2\Delta}} \, \frac{\Gamma\left(\Delta - {i(\hat{\omega}-\hat{\omega}_Q)} \right)}{\Gamma\left(1-\Delta - {i(\hat{\omega}-\hat{\omega}_Q)} \right)}~.
\end{equation}
%\cite{Faulkner:2011tm}
The two parameters $\theta$ and $\omega_{Q}$ are related to the chemical potential and global charge of the operator under a global $U(1)$ symmetry. (If no such symmetry is present, these parameters vanish.) 
%In addition, $T = 1/\pi$, and (for large enough masses of particles in dS$_2$) $2 \omega_Q = \pm \sqrt{4m_{eff}^2-1}$. 
For this model, one can obtain an explicit expression for $\omega_Q$ in terms of the global $U(1)$ charge $q$ of the particular operator, entropy $S$ and charge density $Q$ of the model \cite{Sachdev:2015efa}. For dS$_2$ we identify:
\begin{equation}
\hat{\omega}_Q = \frac{1}{2} \sqrt{4\mu^2-{1}}~, \quad\quad \Delta=\frac{1}{4}~.%\quad\quad T= \frac{1}{\pi}~,%\Delta_\pm = \frac{1}{2} \pm \frac{\sqrt{1-4m^2_{eff}}}{2}~, \quad\quad 
\end{equation}
The structure of (\ref{Gret}) is also found for the correlator of a charged particle in a charged AdS$_2$ black hole \cite{Faulkner:2011tm}. Notice that the poles of (\ref{Gret}) have small real part, similar to the dS ones. In the AdS$_2$ picture, the real part is related to electric repulsion.
% in the AdS$_2$ picture.
%, as in the dS case, a correction to the real part is generated, as a consequence of the electric repulsion.
One could then say that these charged states are less dissipative. It seems dS manages to do the same thing without a (manifest) conserved charge. As stated before, this is reminiscent of glassy/non-ergodic dynamics.

%The parameter $\theta$ is also fixed as:
%\begin{equation}
%e^{\omega_Q/T} = \frac{\sin(\pi \Delta +\theta)}{\sin(\pi \Delta - \theta)}~, \quad\quad Q = \frac{1}{2} - \frac{\theta}{\pi} - \frac{\sin(2\theta)}{4}~.% \quad\quad E = \frac{1}{q} \, \frac{\omega_Q}{2\pi T}~.
%\end{equation}
%:
%The above equation implies that $\theta \to -\theta$ leads to $\hat{\omega}_Q \to -\hat{\omega}_Q$. To match with the dS$_2$ correlator, we would identify $\hat{\omega}_Q = \pm\sqrt{\mu^2-1/4}$ and $T=1/\pi$.

%For $\mu^2 > 1/4$, one of them has $\omega_Q = \sqrt{4\mu^2-1}$ and $\theta$, while the other has $\omega_Q = -\sqrt{4\mu^2-1}$ and $-\theta$. They both have $\Delta = 1/4$. For $\mu^2 < 1/4$, we have $\omega_Q = 0$ for both, and the two $\Delta$'s are given by $\Delta_\pm /2$. Finally, it is worth mentioning that the special case $\mu^2 = 1/4$  gives $\Delta = 1/4$ with $\omega_Q = 0$. 

%For $\mu^2 > 1/4$, one of them has $\omega_Q = \sqrt{4\mu^2-1}$ and $\theta$, while the other has $\omega_Q = -\sqrt{4\mu^2-1}$ and $-\theta$. They both have $\Delta = 1/4$. For $\mu^2 < 1/4$, we have $\omega_Q = 0$ for both, and the two $\Delta$'s are given by $\Delta_\pm /2$. Finally, it is worth mentioning that the special case $\mu^2 = 1/4$  gives $\Delta = 1/4$ with $\omega_Q = 0$. 
Though the similarity of the static patch worldline correlator and the SYK correlators is intriguing, we have not addressed the origin of the product form and why it is local in frequency space. We hope to do so in the near future.

\section{Discussion}

In this section we discuss our results and present an outlook on the nature of the dS static patch.

\subsection{Dissipation in dS worldline theories}

Centaur geometries present a spectrum of quasinormal modes that is notably different from that of AdS black holes. This comparison is relevant. Centaurs exhibit a dS horizon at high temperatures and, as such, are dual to deconfined states in the putative dual quantum mechanics, just like AdS black holes. The centaur spectrum contains modes that are less efficient at thermalizing than those of the AdS$_2$ black hole. In particular there is a collection of quasi-particle like excitations. These correspond to modes that cannot easily access the dS horizon due to scattering from the interpolating region. Their existence is rather surprising from the deconfined plasma perspective. 

There are also dS like modes in the large imaginary region of the complex frequency plane. Up to logarithmic corrections they can be identified with dS horizon physics. A salient feature of these modes, as seen in  (\ref{qnm}) and (\ref{qnm2}), is that for large masses $\mu \gg 1$ they acquire a real part proportional to $\mu$. Therefore, as compared to their energy scale, they are less dissipative than those of a light field. On the other hand, for the AdS$_2$ black hole a heavy particle will become increasingly dissipative. %Once again, from the dual quantum mechanics point of view, it would seem that even highly dissipative modes cannot equilibrate efficiently. 
We draw a parallel to the AdS$_2$ black hole with a constant background $\bold{E}$-field. A charged particle in this spacetime may follow a long lived trajectory due to its electric repulsion from the horizon. Holographically, this reflects the existence of a conserved quantity, which affects the efficiency of equilibration in a thermal state. 

%The poles of the retarded Green's function of a charged particle in AdS$_2$ with an $\bold{E}$-field exhibit a similar real part proportional to the product of the charge and the electric field \cite{Faulkner:2011tm}. We have commented on this interesting connection when comparing the centaur results to those obtained for the SYK model.

Our observations raise a question about the chaotic properties (or lack thereof) of any underlying microscopic model of de Sitter space.\footnote{A localized source of energy sitting at the de Sitter worldline brings in the de Sitter horizon, thus reducing the total horizon entropy, suggesting this is an out of equilibrium configuration. Releasing this source at some early time will eventually backreact due to the blueshift near the horizon, ultimately increasing the size of the horizon and opening up the available space. The effect is opposite to that for ordinary black holes. In the black hole case, releasing such energy leads to an increase in the black hole horizon size which in turn reduces the amount of space outside the horizon. It mimics (somewhat) the discussion of \cite{Gao:2016bin}, in which case a coupling is turned on between the two CFTs of the eternal AdS black hole. Here the coupling seems to be between the original localized mass, and the additional available microstates in the larger de Sitter horizon.} Perhaps these features imply the existence of an emergent entropic conservation law in dS due to disorder. 

%In appendix \ref{geodesics} we discuss timelike geodesics in the centaur geometry. 

%we thus find that the worldline correlator in frequency space is the product of two $SL(2,\mathbb{R})$ invariant correlators, with conformal weight $1/4$. 

\subsection{Worldline holography for higher dimensional dS space-times}

While we have focused on the two dimensional example, it is natural to try to generalize the results of this paper to higher dimensions dS$_2$ $\rightarrow$ dS$_d$.  In this case, the AdS$_2$ region of the geometry would need to be enlarged to AdS$_2$ $\times$ S$^{d-2}$. Notice that this is the natural setup for static patch dS holography.  From the dual quantum mechanics, this is nothing else than the inclusion of a global $SO(d-1)$ symmetry.
% As the number of degrees of freedom is finite, we do not expect that the horizon degrees of freedom can be captured by a quantum field theory. 
The sphere is only realized as the more modest enlargement of the quantum mechanical Hilbert space due to an additional global symmetry. 

Concretely, the proposal would amount to take the higher dimensional static patch (\ref{sp}) and cut it off at some small $r = \epsilon$. At $r = \epsilon$, we could then imagine gluing the geometry to an AdS$_2 \times S^{d-2}_\epsilon$ where $S^{d-2}_\epsilon$ is a small $\epsilon$-sized sphere and the AdS$_2$ is glued to the $(r,t)$ piece of the dS$_d$ static patch geometry (which is the dS$_2$ static patch) at $r=\epsilon$ and has global coordinates in the range $r \in (\epsilon,\infty)$. From this point of view, it would be natural to identify the scale of the sphere to the IR scale which we have also denoted $\epsilon$ in this work. The high temperature regime then corresponds quite naturally to the Euclidean circle being much smaller than the sphere.

Note that this approach is different from previous efforts, such as \cite{Freivogel:2005qh}, to embed dS in higher dimensional AdS/CFT. The crucial difference is that the size of the sphere does not flow under radial evolution in AdS$_2$ evading previous no-go theorems.

\subsection{Centaur dissection}\label{disqg}

To address the question of whether or not there is an autonomous theory of the static patch part of the geometry alone, we would like to cut off the AdS$_2$ half. Morally, this is achieved by adding an energy cutoff to the quantum mechanics, removing high energy states. In doing so, holography suggests we should make the sources dynamical \cite{McGough:2016lol}. This includes the source for the Hamiltonian, namely the worldline metric. Thus, we are led to consider a model of quantum mechanics coupled to worldline gravity. This is quite natural, as in dS there is no decoupling limit for the worldline as opposed to AdS.

%In the following section we consider the structure of correlations directly on the dS$_2$ worldline. These correlations encode the dissipative properties of the near horizon region. %, i.e. the late time behavior of . 

%The number of effective fields is doubled, since corresponding to each operator now also corresponds a fluctuation source.
Another piece of evidence that suggests this direction is the following. Consider the semiclassical limit of Euclidean (bulk) quantum gravity with a positive cosmological constant in $(d+1)$-dimensions. We compute the Euclidean path integral by a saddle point approximation. The dominant saddle is a $(d+1)$-sphere. One of the cycles in the sphere is the Wick rotation of the static patch time coordinate, and its periodicity selects a temperature $T = 1/(2\pi \ell)$. The full path integral, given by integrating over all Euclidean geometries, includes all possible sizes for the thermal circle. In the semiclassical limit $\ell_{pl}/\ell \to 0$, a specific size for the thermal circle is picked. This is a mechanism by which the Euclidean path integral selects a particular temperature for the system in question. This is in sharp contrast to AdS or flat space at finite temperature, where the thermal circle appears as a tunable parameter.
% which must be specified in order to compute the thermal partition function. 

Holographically, are led to consider large $N$ quantum mechanics coupled to a worldline metric $h(\tau)$. On a Euclidean circle, the path integral over the worldline metric becomes an integral of the thermal partition function $Z[\beta]$ over $\beta$. This integral can have cosmological constant $\Lambda$, due to ambiguities in the regularization of the functional determinant \cite{polyakov}. Thus, we end up with:
\begin{equation}
\mathcal{Z}[\Lambda] = \int_{0}^\infty \frac{d\beta}{\beta} \, e^{\Lambda \beta} \, Z_{QM}[\beta]~.
\end{equation}
Notice that the contour of $\beta$ is the non-negative half-line, rather than the imaginary axis we usually consider in thermodynamics. Perhaps for certain large $N$ systems there is a saddle point value for $\beta$. For systems with positive specific heat, the saddle point value of $\beta$ will be a minimum. For systems with negative specific heat it may be a dominant saddle. In the Lorentzian picture, the Hilbert space is restricted to states whose energy is near $\Lambda$. 
\begin{comment}
\red{For instance, consider the propagator in a time-translation invariant theory coupled to a worldline metric. It is some function $G\left(\int_{\tau_1}^{\tau_2}  d\tau \sqrt{h}(\tau) \right) \equiv G(s)$ of the proper time $s$ between the two evaluation points. Path-integrating over the worldline metric, amounts to an integral over $s$, with a possible inclusion of a cosmological constant term:
\begin{equation}
\mathcal{G}(\Lambda) = \int_0^\infty \, ds \, e^{-i \Lambda s} \, \int_{\mathbb{R}} \frac{d\omega}{2\pi} e^{i\omega s} \tilde{G}(\omega) = \int_\mathbb{R} \frac{d\omega}{2\pi i} \, \frac{\tilde{G}(\omega)}{\omega-\Lambda}~.
\end{equation} 
Thus, we obtain the Green function in momentum space at a frequency near $\Lambda$. }
\end{comment}
We will explore this subject in forthcoming work.  

These speculations seem to also be related to the doubling of quasinormal modes in the static patch of de Sitter space as opposed to a single tower of normal modes in global AdS$_2$ \cite{Anninos:2011af}. This is manifest in the expression of the Green functions (\ref{eqds2}) and (\ref{GRd}).

\subsection{On UV universality and weak coupling}

%Lastly, let us comment on a feature of quantum mechanical systems that was emphasized in section \ref{UVQM}, the UV universality of quantum mechanics, and relate it to the particular realization of this phenomenon in this dilaton model.

In section \ref{UVQM} we noted that quantum mechanics does not show a rich landscape of UV structures. It is typically not possible to construct irrelevant operators. Therefore, up to some marginal deformations (as in the DFF model) any quantum mechanical system has a UV free fixed point. When the Hilbert space is finite, the theory may become completely trivial in the UV. From a holographic point of view one would associate this universality to some form of equivalence principle. 
%A region of space-time satisfying the Einstein equations will typically be smooth around any bulk point. 
The physics deep in the neighborhood of a worldline in the holographic bulk associated to the UV regime of the QM, is somehow universal. Given a dual quantum mechanics that is UV free, short-range physics around such a bulk worldline, though universal, will be outside the regime of validity of classical gravity. An exception to the situation above is a boundary for space-time. Constructing a boundary requires turning on a marginal deformation in the dual (like in DFF). 

In our dilaton model, there is an AdS$_2$ region allowing us to extend the validity of the gravity approximation in the bulk. 
%This corresponds to a strongly coupled region of the quantum mechanical model in the UV, once a marginal deformation is turned on (just like in DFF). 
%In our dilaton model, 
The centaur geometry shows a singularity in the UV at $\phi=\phi_0$, signaling that the dual QM becomes weakly coupled in the deep UV. This is quite interesting from the point of view of model construction. From a practical point of view, in our holographic construction we have confined the analysis to the gravitational regime dual to the strong coupling QM region by restricting $\phi < \phi_b \ll \phi_0$.

%\red{DIO :CLOCKS / DEGREES OF FREEDOM IN THE DILATON POTENTIAL?}

\subsection{Future directions}

The clear outstanding challenge is to build a quantum mechanical model dual to the geometric description outlined in this paper.

Given the recent developments in understanding the AdS$_2$ holographic dual of the SYK model  \cite{Sachdev:1992fk,pg,kitaev,Sachdev:2015efa,Anninos:2013nra,Anninos:2016szt,Polchinski:2016xgd,Maldacena:2016hyu,Maldacena:2016upp}, one approach may lie in finding an appropriate relevant deformation that gives rise to an infrared dS geometry. What does that mean in the context of this model? A working definition of \textit{de Sitterness} consists in finding a quasinormal spectrum with the properties described in section \ref{quasiwaves}. In particular we expect the presence of a tower of equally spaced almost imaginary quasi-normal modes, not only in the region $\omega \ell  \gg T$ but also at $\omega \ell \gtrsim T$. While this finite temperature solution must correspond to a deconfined phase, we expect the physics to be less dissipative than the AdS$_2$ black hole regime. Relatedly, we would like to reproduce the doubling structure of the Green function in momentum space (\ref{eqds2}). The doubling structure is not present for the underformed SYK constructions. Perhaps it is connected to the second boundary in global AdS$_2$,
%\red{This is can be conformally mapped to the causal structure of global dS.}
which is a consequence of radial quantization in $(0+1)$-dimensions. The state operator mapping relates operator insertions to states in a double copy of the quantum mechanics \cite{Sen:2011cn}. Removing the second boundary might require coupling one of the copies to $(0+1)$-dimensional quantum gravity. As discussed in \ref{disqg}, this approach may be necessary to isolate the dS geometry from the AdS$_2$ boundary.
% discussed in this work.

%The situation is not that straightforward, however. 
However, there is a subtlety to be overcome. Centaur geometries with a stable dS interior required a dilaton $\Phi=\phi_0 -\phi$ with a deviation $\phi >0$ from the value fixing the large ground state degeneracy. This is opposite to the situation in the SYK setup, where the dilaton grows toward the AdS$_2$ boundary \cite{Maldacena:2016upp}.\footnote{More closely related to SYK would be the situation discussed in appendix \ref{negdil}.} It leads to the presence of a weakly coupled region of the quantum mechanics at the boundary of the geometry and not deep inside the bulk. The consequences for the density of states of this theory remain to be understood.
%, as in the usual case

These comments hint at a new more general framework for quantum mechanical systems coupled to gravity in the context of holography.

\section*{Acknowledgements}

It was a great pleasure discussing this work with Tarek Anous, Tom Banks, Frederik Denef, Sean Hartnoll, Ben Freivogel, Juan Maldacena, Aron Wall, Nabil Iqbal, Douglas Stanford and especially Nati Seiberg. D.A. is funded by the AMIAS and the NSF. This project has received funding from the European Research Council (ERC) under the European Union’s Horizon 2020 research and innovation programme (grant agreement No 715656).

%\newpage

\appendix

\section{Dimensional reduction of Einstein gravity}\label{nariaisec}

We consider the effective two-dimensional dilaton gravity model obtained by reducing the cosmological Einstein action on a two-sphere. The four-dimensional metric takes the form
\begin{equation}
ds^2 = \frac{1}{\sqrt{\phi}} g_{\mu\nu} dx^\mu dx^\nu + 4\phi d\Omega^2~.
\end{equation}
The two-dimensional theory is:
\begin{equation}\label{nariai}
S_L = \int d^2x \sqrt{-g} \left( \phi R + \frac{1}{2\sqrt{\phi}}  - \frac{3s}{2} \sqrt{\phi} \right)~. %  \partial_\mu \phi \partial^\mu \phi +
\end{equation}
where the two-sphere area is $16\pi \phi$ and we have rescaled the four-dimensional cosmological constant for convenience. Positive cosmological constant corresponds to $s=+1$ and negative to $s=-1$. The equations of motion are given by:
\begin{eqnarray}
\nabla_{\mu} \nabla_{\nu} \phi - g_{\mu\nu} \nabla^2 \phi + \frac{1}{2} g_{\mu\nu} V(\phi) = 0~, \\
R + V'(\phi) = 0~,
\end{eqnarray}
where $V(\phi) = (1/2\sqrt{\phi} - 3 s \sqrt{\phi}/2)$. Notice that for $s = +1$ there is a solution with a constant $\phi = 1/3$. At this special point the two-dimensional metric is dS$_2$. This is the so called Nariai solution. There is no such solution for $s=-1$. 

There is another solution with a running dilaton. Using a B\"{a}cklund transformation, it can be written as:
\begin{equation}
ds^2 = -dT^2  N(\phi)  + \frac{d\phi^2}{ N(\phi) }~,
\end{equation}
where $N(\phi) = \left( \sqrt{\phi} - s \, \phi^{3/2} \right)$. This is the ordinary dS$_4$ static patch. The more general solution will be the dimensional reduction of the Schwarszchild-de Sitter geometry. The dilaton will be monotonic and hence decrease toward the black hole horizon and increase toward the de Sitter horizon. 

%For a more general potential, we can have a different situation. Take for example $V(\phi) = \sqrt{\phi^2+\epsilon}$ with $c$ a small positive number. Then we have:
%\begin{equation}
%N(\phi) =  \frac{1}{2} \left(\phi \sqrt{\epsilon+\phi^2}+\epsilon \log \left(\sqrt{\epsilon+\phi^2}+\phi\right)\right)~,
%\end{equation}
%a function well approximated by $\text{sign} (\phi) \, \phi^2$ for small $\epsilon$. In other words the two-dimensional metric looks close to dS$_2$ for negative $\phi$ and close to AdS$_2$ for positive $\phi$. It is a centaur geometry. 
%
%\subsection{AdS$_2 \times S^2$ dilaton gravity}
%
%We can also consider zero cosmological constant but with a Maxwell field. The dimensionally reduced action is now:
%\begin{equation}
%S_L = \int dt \sqrt{-g}  \left( \phi R + \frac{1}{2\sqrt{\phi}}  - {\phi}^{3/2} F_{\mu\nu} F^{\mu\nu} \right)~.
%\end{equation}
%A solution with $\phi$ constant and constant electric flux through the two-sphere gives an effectively negative cosmological constant and an AdS$_2 \times S^2$ solution. 
%
\section{Negative boundary values for the dilaton}\label{negdil}

We give some details about the solutions to the dilaton theory (\ref{dilgrav}) with  a dilaton that is  negative near the AdS$_2$ boundary. For this theory we take $V(\phi) =  - 2  \left( \sqrt{\phi^2+\epsilon^2}-\epsilon \right)$. The solution is given again by:
\begin{equation}
ds^2 = -dt^2  N(\phi) + \frac{d\phi^2}{\ell^2 N(\phi)}~, 
\end{equation}
with $N(\phi) = \int^{\phi_h}_\phi dz V(z)$, with $\phi_b \le \phi< \phi_h$.  
%Consider first $c^2 > 0$. The range of $\phi$ is now $\phi \in (\phi_h,-\phi_b)$. Here $\phi_b >0$ is the value of the dilaton near the AdS$_2$ boundary and $\phi_h \approx c > 0$ for $c\gg \epsilon$. For $c^2 = -\gamma^2 < 0$, we have non-centaur solution (\ref{noncentaur}), for which $\phi \in (-\phi_h,-\phi_b)$. 
Computing the thermodynamics as in the main text, we now find that the specific heat of the dS$_2$ like solution with $\phi_h>0$ is negative, whereas that for $\phi_h<0$ is positive. %Also the entropy is now decreasing away from the topological value $S_0 = \phi_0/4G$.
Essentially, the Euclidean action flips sign as compared to the analysis of the main text. 

Something analogous happens for solutions that occur if we deform away from the constant dilaton dS$_2$ vacua of the dilaton action (\ref{nariai}). Then, one of the dS$_2$ horizons becomes a black hole horizon, with negative specific heat, whereas the other becomes the de Sitter cosmological horizon, with positive specific heat. The one with negative specific heat has a decreasing horizon size. We associate the case studied in the main text, i.e. the centaur geometry with positive dilaton at the AdS$_2$ boundary, with the part of the Nariai geometry between the worldline and the de Sitter cosmological horizon. The Nariai geometry between the worldline and the black hole horizon is associated to the centaur geometry with negative dilaton at the AdS$_2$ boundary. (See \cite{Kyono:2017jtc} for a different appearance of dS$_2$ in dilaton gravity in a related setup.)

\section{Expression for $\gamma({\hat{\omega}})$, $\beta({\hat{\omega}})$ and $\sigma(\hat{\omega})$}\label{appendixsigma}
 
We present the explicit expressions for $\gamma({\hat{\omega}})$ and $\beta({\hat{\omega}})$:
\begin{multline}
\gamma({\hat{\omega}}) = 
\frac{2^{i \hat{\omega}}}{2^{\hat{\omega}}} \left(\frac{i \cos \left(\frac{1}{2} \pi  (\Delta +i \hat{\omega} )\right) \Gamma \left(\frac{1}{2} (\Delta +i \hat{\omega} +1)\right) \cos \left(\frac{1}{2} \pi  (\tilde{\Delta}+\hat{\omega} )\right) \Gamma \left(\frac{\tilde{\Delta}-\hat{\omega} }{2}\right)}{\Gamma \left(\frac{1}{2} (\Delta -i \hat{\omega} )\right) \Gamma \left(\frac{1}{2} (\tilde{\Delta}+\hat{\omega} +1)\right)}\right. \\ \left.
+\frac{\sin \left(\frac{1}{2} \pi  (\Delta +i \hat{\omega} )\right) \Gamma \left(\frac{1}{2} (\Delta +i \hat{\omega} )\right) \sin \left(\frac{1}{2} \pi  (\tilde{\Delta}+\hat{\omega} )\right) \Gamma \left(\frac{1}{2} (\tilde{\Delta}-\hat{\omega} +1)\right)}{\Gamma \left(\frac{1}{2} (\Delta -i \hat{\omega} +1)\right) \Gamma \left(\frac{\tilde{\Delta}+\hat{\omega} }{2}\right)}\right)~,
\end{multline}
and
\begin{multline}
\beta({\hat{\omega}}) = \frac{2}{2^{(1-i) \hat{\omega}}\pi}  \left(\frac{\sin \left(\frac{1}{2} \pi  (\Delta +i \hat{\omega} )\right) \Gamma \left(\frac{1}{2} (\Delta +i \hat{\omega} )\right) \cos \left(\frac{1}{2} \pi  (\tilde{\Delta}+\hat{\omega} )\right) \Gamma \left(\frac{1}{2} (\tilde{\Delta}-\hat{\omega} +1)\right)}{\Gamma \left(\frac{1}{2} (\Delta -i \hat{\omega} +1)\right) \Gamma \left(\frac{\tilde{\Delta}+\hat{\omega} }{2}\right)} \right. \\ \left. 
 -\frac{i \cos \left(\frac{1}{2} \pi  (\Delta +i \hat{\omega} )\right) \Gamma \left(\frac{1}{2} (\Delta +i \hat{\omega} +1)\right) \sin \left(\frac{1}{2} \pi  (\tilde{\Delta}+\hat{\omega} )\right) \Gamma \left(\frac{\tilde{\Delta}-\hat{\omega} }{2}\right)}{\Gamma \left(\frac{1}{2} (\Delta -i \hat{\omega} )\right) \Gamma \left(\frac{1}{2} (\tilde{\Delta}+\hat{\omega} +1)\right)}\right)~.
\end{multline}
To derive the above relations we have expressed the associated Legendre functions and their first derivatives at the origin in terms of Gamma-functions. For example, 
\begin{equation}
P^q_p(0) = \frac{2^{q } \cos \left(\frac{1}{2} \pi  (q +p )\right) \Gamma \left(\frac{1}{2} (q +p +1)\right)}{\sqrt{\pi } \Gamma \left(\frac{1}{2} (-q +p +2)\right)}~.
\end{equation} 
Finally, we express $\sigma(\hat{\omega})$ as a function of $\beta(\hat{\omega})$ and $\gamma(\hat{\omega})$. We have $\sigma(\hat{\omega}) = N(\hat{\omega})/D(\hat{\omega})$ where:
%Expressed as numerator over denominator ($\nu = \tilde{\Delta}-1$):
\begin{equation}
N(\hat{\omega}) = \frac{ -4 \gamma(\hat{\omega}) -\pi  \beta(\hat{\omega})  e^{-\pi  (\hat{\omega} +i \tilde{\Delta} )} \left(e^{2 i \pi   \tilde{\Delta} }+e^{2 \pi  \hat{\omega} }+2\right) \csc (\pi  ( \tilde{\Delta} +i \hat{\omega} )}{2 \Gamma (1-\tilde{\Delta} -i \hat{\omega} )\sqrt{\pi }}~,
% \gamma  \Gamma \left(-\nu -\frac{1}{2}\right)-\sqrt{\pi } \beta  2^{2\nu } e^{-\pi  (\omega +i \nu )} \left(e^{2 i \pi  \nu }+e^{2 \pi  \omega }+2\right) \sin (\pi  \nu ) \Gamma (-2 \nu -1) \Gamma (\nu +1) \csc (\pi  (\nu +i \omega ))
\end{equation}
\begin{equation}
D(\hat{\omega})  = \frac{- \gamma(\hat{\omega})  4^{ \tilde{\Delta}-1} \csc (\pi \tilde{\Delta} ) \Gamma \left(\tilde{\Delta} - \frac{1}{2}\right)-i \sqrt{\pi } \beta(\hat{\omega})  \Gamma (2-\tilde{\Delta} ) \Gamma (2 ( \tilde{\Delta}-1) )}{4^{\tilde{\Delta}-1}\Gamma (-2  \tilde{\Delta} + 1) \Gamma (\tilde{\Delta})\Gamma (\tilde{\Delta} -i \hat{\omega} )}~.
%\frac{2 \Gamma (-\nu -i \omega )}{\Gamma (\nu -i \omega +1)\sin (\pi  \nu ) \Gamma (-2 \nu -1) \Gamma (\nu +1)} \left(\gamma  2^{2\nu } \Gamma \left(\nu +\frac{1}{2}\right)-i \sqrt{\pi } \beta  \sin (\pi  \nu ) \Gamma (1-\nu ) \Gamma (2 \nu )\right)
\end{equation}


\begin{thebibliography}{1}

%\cite{Maldacena:1997re}
\bibitem{Maldacena:1997re} 
  J.~M.~Maldacena,
  ``The Large N limit of superconformal field theories and supergravity,''
  Int.\ J.\ Theor.\ Phys.\  {\bf 38}, 1113 (1999)
  [Adv.\ Theor.\ Math.\ Phys.\  {\bf 2}, 231 (1998)]
  doi:10.1023/A:1026654312961
  [hep-th/9711200].
  %%CITATION = doi:10.1023/A:1026654312961;%%
  %12176 citations counted in INSPIRE as of 12 Oct 2016

%\cite{Sachdev:1992fk}
\bibitem{Sachdev:1992fk} 
  S.~Sachdev and J.~Ye,
  ``Gapless spin fluid ground state in a random, quantum Heisenberg magnet,''
  Phys.\ Rev.\ Lett.\  {\bf 70}, 3339 (1993)
  doi:10.1103/PhysRevLett.70.3339
  [cond-mat/9212030].
  %%CITATION = doi:10.1103/PhysRevLett.70.3339;%%
  %53 citations counted in INSPIRE as of 12 Oct 2016

\bibitem{pg}
O. Parcollet and A. Georges. Non-Fermi-liquid regime of a doped Mott insulator - 1999. Phys.Rev.,B59,5341 cond-mat/9806119

\bibitem{kitaev}
A. Kitaev. A simple model of quantum holography - 2015. KITP strings seminar and Entanglement program (Feb. 12, April 7, and May 27,). http://online.kitp.ucsb.edu/online/entangled15

%\cite{Sachdev:2015efa}
\bibitem{Sachdev:2015efa} 
  S.~Sachdev,
  ``Bekenstein-Hawking Entropy and Strange Metals,''
  Phys.\ Rev.\ X {\bf 5}, no. 4, 041025 (2015)
  doi:10.1103/PhysRevX.5.041025
  [arXiv:1506.05111 [hep-th]].
  %%CITATION = doi:10.1103/PhysRevX.5.041025;%%
  %32 citations counted in INSPIRE as of 12 Oct 2016
  
      %\cite{Anninos:2013nra}
\bibitem{Anninos:2013nra} 
  D.~Anninos, T.~Anous, P.~de Lange and G.~Konstantinidis,
  ``Conformal quivers and melting molecules,''
  JHEP {\bf 1503}, 066 (2015)
  %doi:10.1007/JHEP03(2015)066
  [arXiv:1310.7929 [hep-th]].
  %%CITATION = doi:10.1007/JHEP03(2015)066;%%
  %5 citations counted in INSPIRE as of 02 Oct 2016
  
  %\cite{Anninos:2016szt}
\bibitem{Anninos:2016szt} 
  D.~Anninos, T.~Anous and F.~Denef,
  ``Disordered Quivers and Cold Horizons,''
  arXiv:1603.00453 [hep-th].
  %%CITATION = ARXIV:1603.00453;%%
  %7 citations counted in INSPIRE as of 02 Oct 2016
  
  
%\cite{Polchinski:2016xgd}\cite{Maldacena:2016hyu}\cite{Maldacena:2016upp}
\bibitem{Polchinski:2016xgd} 
  J.~Polchinski and V.~Rosenhaus,
  ``The Spectrum in the Sachdev-Ye-Kitaev Model,''
  JHEP {\bf 1604}, 001 (2016)
  doi:10.1007/JHEP04(2016)001
  [arXiv:1601.06768 [hep-th]].
  %%CITATION = doi:10.1007/JHEP04(2016)001;%%
  %27 citations counted in INSPIRE as of 12 Oct 2016
  
  %\cite{Maldacena:2016hyu}\cite{Maldacena:2016upp}
\bibitem{Maldacena:2016hyu} 
  J.~Maldacena and D.~Stanford,
  ``Comments on the Sachdev-Ye-Kitaev model,''
  arXiv:1604.07818 [hep-th].
  %%CITATION = ARXIV:1604.07818;%%
  %26 citations counted in INSPIRE as of 12 Oct 2016
  
  %\cite{Maldacena:2016upp}
\bibitem{Maldacena:2016upp} 
  J.~Maldacena, D.~Stanford and Z.~Yang,
  ``Conformal symmetry and its breaking in two dimensional Nearly Anti-de-Sitter space,''
  arXiv:1606.01857 [hep-th].
  %%CITATION = ARXIV:1606.01857;%%
  %15 citations counted in INSPIRE as of 02 Oct 2016

  \bibitem{Hartnoll:2009sz} 
  S.~A.~Hartnoll,
  ``Lectures on holographic methods for condensed matter physics,''
  Class.\ Quant.\ Grav.\  {\bf 26}, 224002 (2009)
  doi:10.1088/0264-9381/26/22/224002
  [arXiv:0903.3246 [hep-th]].
  %%CITATION = doi:10.1088/0264-9381/26/22/224002;%%
  %965 citations counted in INSPIRE as of 12 Oct 2016
  
  %\cite{Polchinski:2000uf}
\bibitem{Polchinski:2000uf} 
  J.~Polchinski and M.~J.~Strassler,
  %``The String dual of a confining four-dimensional gauge theory,''
  hep-th/0003136.
  %%CITATION = HEP-TH/0003136;%%
  %563 citations counted in INSPIRE as of 07 Nov 2016
  
  %\cite{Majumdar:1947eu}
\bibitem{Majumdar:1947eu} 
  S.~D.~Majumdar,
  ``A class of exact solutions of Einstein's field equations,''
  Phys.\ Rev.\  {\bf 72}, 390 (1947).
  doi:10.1103/PhysRev.72.390
  %%CITATION = doi:10.1103/PhysRev.72.390;%%
  %313 citations counted in INSPIRE as of 03 Oct 2016
  
\bibitem{papapetrou}  
  A. Papapetrou,
  ``Static solution of the equations of the gravitational field for an arbitrary charge distribution,"  
  Proc. Roy. Irish Acad., A51, 191 (1947)

  %\cite{Parikh:2004wh}\cite{Banks:2012ic}\cite{Dong:2010pm}\cite{Goheer:2002vf}\cite{Anninos:2012qw}
\bibitem{Parikh:2004wh} 
  M.~K.~Parikh and E.~P.~Verlinde,
  ``De Sitter holography with a finite number of states,''
  JHEP {\bf 0501}, 054 (2005)
  doi:10.1088/1126-6708/2005/01/054
  [hep-th/0410227].
  %%CITATION = doi:10.1088/1126-6708/2005/01/054;%%
  %35 citations counted in INSPIRE as of 12 Oct 2016
  
  %\cite{Gibbons:1977mu}
\bibitem{Gibbons:1977mu} 
  G.~W.~Gibbons and S.~W.~Hawking,
  ``Cosmological Event Horizons, Thermodynamics, and Particle Creation,''
  Phys.\ Rev.\ D {\bf 15}, 2738 (1977).
  %%CITATION = PHRVA,D15,2738;%%
  %1438 citations counted in INSPIRE as of 29 Mar 2014
  
%\cite{Banks:2012ic}
\bibitem{Banks:2012ic} 
  T.~Banks,
  ``Holographic spacetime,''
  Int.\ J.\ Mod.\ Phys.\ D {\bf 21}, 1241004 (2012).
  doi:10.1142/S0218271812410040
  %%CITATION = doi:10.1142/S0218271812410040;%%
  %3 citations counted in INSPIRE as of 12 Oct 2016
  
  %\cite{Dong:2010pm}
\bibitem{Dong:2010pm} 
  X.~Dong, B.~Horn, E.~Silverstein and G.~Torroba,
  ``Micromanaging de Sitter holography,''
  Class.\ Quant.\ Grav.\  {\bf 27}, 245020 (2010)
  doi:10.1088/0264-9381/27/24/245020
  [arXiv:1005.5403 [hep-th]].
  %%CITATION = doi:10.1088/0264-9381/27/24/245020;%%
  %48 citations counted in INSPIRE as of 12 Oct 2016
  
  %\cite{Goheer:2002vf}
\bibitem{Goheer:2002vf} 
  N.~Goheer, M.~Kleban and L.~Susskind,
  ``The Trouble with de Sitter space,''
  JHEP {\bf 0307}, 056 (2003)
  doi:10.1088/1126-6708/2003/07/056
  [hep-th/0212209].
  %%CITATION = doi:10.1088/1126-6708/2003/07/056;%%
  %178 citations counted in INSPIRE as of 12 Oct 2016  
  
    %\cite{Anninos:2012qw}
\bibitem{Anninos:2012qw} 
  D.~Anninos,
  ``De Sitter Musings,''
  Int.\ J.\ Mod.\ Phys.\ A {\bf 27}, 1230013 (2012)
  doi:10.1142/S0217751X1230013X
  [arXiv:1205.3855 [hep-th]].
  %%CITATION = doi:10.1142/S0217751X1230013X;%%
  %47 citations counted in INSPIRE as of 12 Oct 2016
  
  %\cite{Verlinde:2016toy}
\bibitem{Verlinde:2016toy} 
  E.~P.~Verlinde,
  ``Emergent Gravity and the Dark Universe,''
  arXiv:1611.02269 [hep-th].
  %%CITATION = ARXIV:1611.02269;%%
  
  %\cite{Maltz:2016max}
\bibitem{Maltz:2016max} 
  J.~Maltz,
  ``de Sitter Harmonies: Cosmological Spacetimes as Resonances,''
  arXiv:1611.03491 [hep-th].
  %%CITATION = ARXIV:1611.03491;%%
  
    %\cite{Freivogel:2005qh}
\bibitem{Freivogel:2005qh} 
  B.~Freivogel, V.~E.~Hubeny, A.~Maloney, R.~C.~Myers, M.~Rangamani and S.~Shenker,
  ``Inflation in AdS/CFT,''
  JHEP {\bf 0603}, 007 (2006)
  doi:10.1088/1126-6708/2006/03/007
  [hep-th/0510046].
  %%CITATION = doi:10.1088/1126-6708/2006/03/007;%%
  %100 citations counted in INSPIRE as of 12 Oct 2016
  
  %\cite{Lowe:2010np}
\bibitem{Lowe:2010np} 
  D.~A.~Lowe and S.~Roy,
  ``Punctuated eternal inflation via AdS/CFT,''
  Phys.\ Rev.\ D {\bf 82}, 063508 (2010)
  doi:10.1103/PhysRevD.82.063508
  [arXiv:1004.1402 [hep-th]].
  %%CITATION = doi:10.1103/PhysRevD.82.063508;%%
  %15 citations counted in INSPIRE as of 27 Dec 2016
  
  %\cite{Banks:1996vh}\cite{Itzhaki:1998dd}\cite{deAlfaro:1976vlx}\cite{Strominger:1998yg}\cite{Sen:2011cn}\cite{Majumdar:1947eu}\cite{papapetrou}
\bibitem{Banks:1996vh} 
  T.~Banks, W.~Fischler, S.~H.~Shenker and L.~Susskind,
  ``M theory as a matrix model: A Conjecture,''
  Phys.\ Rev.\ D {\bf 55}, 5112 (1997)
  %doi:10.1103/PhysRevD.55.5112
  [hep-th/9610043].
  %%CITATION = doi:10.1103/PhysRevD.55.5112;%%
  %2543 citations counted in INSPIRE as of 02 Oct 2016
  

 
%\cite{Itzhaki:1998dd}
\bibitem{Itzhaki:1998dd} 
  N.~Itzhaki, J.~M.~Maldacena, J.~Sonnenschein and S.~Yankielowicz,
  ``Supergravity and the large N limit of theories with sixteen supercharges,''
  Phys.\ Rev.\ D {\bf 58}, 046004 (1998)
  %doi:10.1103/PhysRevD.58.046004
  [hep-th/9802042].
  %%CITATION = doi:10.1103/PhysRevD.58.046004;%%
  %834 citations counted in INSPIRE as of 02 Oct 2016
  
  %\cite{deAlfaro:1976vlx}
\bibitem{deAlfaro:1976vlx} 
  V.~de Alfaro, S.~Fubini and G.~Furlan,
  ``Conformal Invariance in Quantum Mechanics,''
  Nuovo Cim.\ A {\bf 34}, 569 (1976).
  doi:10.1007/BF02785666
  %%CITATION = doi:10.1007/BF02785666;%%
  %394 citations counted in INSPIRE as of 02 Oct 2016
  
    %\cite{Hammer:2005sa}
\bibitem{Hammer:2005sa} 
  H.-W.~Hammer and B.~G.~Swingle,
  %``On the limit cycle for the 1/r**2 potential in momentum space,''
  Annals Phys.\  {\bf 321}, 306 (2006)
  doi:10.1016/j.aop.2005.04.017
  [quant-ph/0503074].
  %%CITATION = doi:10.1016/j.aop.2005.04.017;%%
  %20 citations counted in INSPIRE as of 07 Mar 2017
  
    %\cite{Anninos:2015eji}\cite{Anninos:2009yc}\cite{Anninos:2010gh}\cite{Anninos:2011vd}
\bibitem{Anninos:2015eji} 
  D.~Anninos, F.~Denef and R.~Monten,
  ``Grassmann Matrix Quantum Mechanics,''
  JHEP {\bf 1604}, 138 (2016)
  doi:10.1007/JHEP04(2016)138
  [arXiv:1512.03803 [hep-th]].
  %%CITATION = doi:10.1007/JHEP04(2016)138;%%
  %2 citations counted in INSPIRE as of 12 Oct 2016
  
  %\cite{Anninos:2016klf}
\bibitem{Anninos:2016klf} 
  D.~Anninos and G.~A.~Silva,
  ``Solvable Quantum Grassmann Matrices,''
  arXiv:1612.03795 [hep-th].
  %%CITATION = ARXIV:1612.03795;%%
  %6 citations counted in INSPIRE as of 09 Mar 2017


    \bibitem{upcoming}
 D.~Anninos and D.~Hofman, 
 ``Circle Saddles,"
 in preparation



  %\cite{Anninos:2009jt}
\bibitem{Anninos:2009jt} 
  D.~Anninos,
  ``Sailing from Warped AdS(3) to Warped dS(3) in Topologically Massive Gravity,''
  JHEP {\bf 1002}, 046 (2010)
  doi:10.1007/JHEP02(2010)046
  [arXiv:0906.1819 [hep-th]].
  %%CITATION = doi:10.1007/JHEP02(2010)046;%%
  %28 citations counted in INSPIRE as of 12 Mar 2017

  
%\cite{Anninos:2009yc}
\bibitem{Anninos:2009yc} 
  D.~Anninos and T.~Hartman,
  ``Holography at an Extremal De Sitter Horizon,''
  JHEP {\bf 1003}, 096 (2010)
  doi:10.1007/JHEP03(2010)096
  [arXiv:0910.4587 [hep-th]].
  %%CITATION = doi:10.1007/JHEP03(2010)096;%%
  %30 citations counted in INSPIRE as of 12 Oct 2016
  
  %\cite{Anninos:2010gh}
\bibitem{Anninos:2010gh} 
  D.~Anninos and T.~Anous,
  ``A de Sitter Hoedown,''
  JHEP {\bf 1008}, 131 (2010)
  doi:10.1007/JHEP08(2010)131
  [arXiv:1002.1717 [hep-th]].
  %%CITATION = doi:10.1007/JHEP08(2010)131;%%
  %25 citations counted in INSPIRE as of 12 Oct 2016
  
  %\cite{Anninos:2011vd}
\bibitem{Anninos:2011vd} 
  D.~Anninos, S.~de Buyl and S.~Detournay,
  ``Holography For a De Sitter-Esque Geometry,''
  JHEP {\bf 1105}, 003 (2011)
  doi:10.1007/JHEP05(2011)003
  [arXiv:1102.3178 [hep-th]].
  %%CITATION = doi:10.1007/JHEP05(2011)003;%%
  %17 citations counted in INSPIRE as of 12 Oct 2016

    %\cite{Strominger:1998yg}
\bibitem{Strominger:1998yg} 
  A.~Strominger,
  ``AdS(2) quantum gravity and string theory,''
  JHEP {\bf 9901}, 007 (1999)
  doi:10.1088/1126-6708/1999/01/007
  [hep-th/9809027].
  %%CITATION = doi:10.1088/1126-6708/1999/01/007;%%
  %212 citations counted in INSPIRE as of 12 Oct 2016
  
  %\cite{Sen:2011cn}
\bibitem{Sen:2011cn} 
  A.~Sen,
  ``State Operator Correspondence and Entanglement in $AdS_2/CFT_1$,''
  Entropy {\bf 13}, 1305 (2011)
  %doi:10.3390/e13071305
  [arXiv:1101.4254 [hep-th]].
  %%CITATION = doi:10.3390/e13071305;%%
  %51 citations counted in INSPIRE as of 02 Oct 2016
  
  
    %\cite{Strominger:2003tm,verlinde}
\bibitem{Strominger:2003tm} 
  A.~Strominger,
  ``A Matrix model for AdS(2),''
  JHEP {\bf 0403}, 066 (2004)
  doi:10.1088/1126-6708/2004/03/066
  [hep-th/0312194].
  %%CITATION = doi:10.1088/1126-6708/2004/03/066;%%
  %42 citations counted in INSPIRE as of 08 Nov 2016
  
  \bibitem{verlinde}
  H.~Verlinde,
  ``Superstrings on AdS$_2$ and superconformal matrix quantum mechanics,"
  [hep-th/0403024]


  

  
  %\cite{Cavaglia:1998xj}\cite{Grumiller:2007ju}
\bibitem{Cavaglia:1998xj} 
  M.~Cavaglia,
  ``Geometrodynamical formulation of two-dimensional dilaton gravity,''
  Phys.\ Rev.\ D {\bf 59}, 084011 (1999)
  %doi:10.1103/PhysRevD.59.084011
  [hep-th/9811059].
  %%CITATION = doi:10.1103/PhysRevD.59.084011;%%
  %21 citations counted in INSPIRE as of 02 Oct 2016
  
    %\cite{Witten:1998zw}
\bibitem{Witten:1998zw} 
  E.~Witten,
  ``Anti-de Sitter space, thermal phase transition, and confinement in gauge theories,''
  Adv.\ Theor.\ Math.\ Phys.\  {\bf 2}, 505 (1998)
  [hep-th/9803131].
  %%CITATION = HEP-TH/9803131;%%
  %2492 citations counted in INSPIRE as of 27 Dec 2016
  
  %\cite{Grumiller:2007ju}
\bibitem{Grumiller:2007ju} 
  D.~Grumiller and R.~McNees,
  ``Thermodynamics of black holes in two (and higher) dimensions,''
  JHEP {\bf 0704}, 074 (2007)
  %doi:10.1088/1126-6708/2007/04/074
  [hep-th/0703230 [HEP-TH]].
  %%CITATION = doi:10.1088/1126-6708/2007/04/074;%%
  %53 citations counted in INSPIRE as of 02 Oct 2016
  
%\cite{Gegenberg:1994pv}
\bibitem{Gegenberg:1994pv} 
  J.~Gegenberg, G.~Kunstatter and D.~Louis-Martinez,
  ``Observables for two-dimensional black holes,''
  Phys.\ Rev.\ D {\bf 51}, 1781 (1995)
  doi:10.1103/PhysRevD.51.1781
  [gr-qc/9408015].
  %%CITATION = doi:10.1103/PhysRevD.51.1781;%%
  %124 citations counted in INSPIRE as of 07 Nov 2017
  
  %\cite{Grumiller:2006rc}
\bibitem{Grumiller:2006rc} 
  D.~Grumiller and R.~Meyer,
  ``Ramifications of lineland,''
  Turk.\ J.\ Phys.\  {\bf 30}, 349 (2006)
  [hep-th/0604049].
  %%CITATION = HEP-TH/0604049;%%
  %47 citations counted in INSPIRE as of 07 Nov 2017
  
  
  


%%\cite{Hamilton:2005ju}
%\bibitem{Hamilton:2005ju} 
%  A.~Hamilton, D.~N.~Kabat, G.~Lifschytz and D.~A.~Lowe,
%  ``Local bulk operators in AdS/CFT: A Boundary view of horizons and locality,''
%  Phys.\ Rev.\ D {\bf 73}, 086003 (2006)
%  doi:10.1103/PhysRevD.73.086003
%  [hep-th/0506118].
%  %%CITATION = doi:10.1103/PhysRevD.73.086003;%%
%  %105 citations counted in INSPIRE as of 11 Dec 2016


  
%\cite{LopezOrtega:2006my}\cite{Cardoso:2003sw}\cite{Anninos:2011af}\cite{Faulkner:2011tm}\cite{Freivogel:2005qh}
\bibitem{LopezOrtega:2006my} 
  A.~Lopez-Ortega,
  ``Quasinormal modes of D-dimensional de Sitter spacetime,''
  Gen.\ Rel.\ Grav.\  {\bf 38}, 1565 (2006)
  doi:10.1007/s10714-006-0335-9
  [gr-qc/0605027].
  %%CITATION = doi:10.1007/s10714-006-0335-9;%%
  %38 citations counted in INSPIRE as of 12 Oct 2016
  
%\cite{Cardoso:2003sw}
\bibitem{Cardoso:2003sw} 
  V.~Cardoso and J.~P.~S.~Lemos,
  ``Quasinormal modes of the near extremal Schwarzschild-de Sitter black hole,''
  Phys.\ Rev.\ D {\bf 67}, 084020 (2003)
  doi:10.1103/PhysRevD.67.084020
  [gr-qc/0301078].
  %%CITATION = doi:10.1103/PhysRevD.67.084020;%%
  %137 citations counted in INSPIRE as of 12 Oct 2016
  
  
  %\cite{Anninos:2011af}
\bibitem{Anninos:2011af} 
  D.~Anninos, S.~A.~Hartnoll and D.~M.~Hofman,
  ``Static Patch Solipsism: Conformal Symmetry of the de Sitter Worldline,''
  Class.\ Quant.\ Grav.\  {\bf 29}, 075002 (2012)
  %doi:10.1088/0264-9381/29/7/075002
  [arXiv:1109.4942 [hep-th]].
  %%CITATION = doi:10.1088/0264-9381/29/7/075002;%%
  %35 citations counted in INSPIRE as of 02 Oct 2016
  
    %\cite{Faulkner:2011tm}
\bibitem{Faulkner:2011tm} 
  T.~Faulkner, N.~Iqbal, H.~Liu, J.~McGreevy and D.~Vegh,
  ``Holographic non-Fermi liquid fixed points,''
  Phil.\ Trans.\ Roy.\ Soc.\ A {\bf  369}, 1640 (2011)
  %doi:10.1098/rsta.2010.0354
  [arXiv:1101.0597 [hep-th]].
  %%CITATION = doi:10.1098/rsta.2010.0354;%%
  %57 citations counted in INSPIRE as of 02 Oct 2016
  

%\cite{Gao:2016bin}
\bibitem{Gao:2016bin} 
  P.~Gao, D.~L.~Jafferis and A.~Wall,
  ``Traversable Wormholes via a Double Trace Deformation,''
  arXiv:1608.05687 [hep-th].
  %%CITATION = ARXIV:1608.05687;%%
  %1 citations counted in INSPIRE as of 14 Oct 2016
  

  

  
  %\cite{McGough:2016lol}
\bibitem{McGough:2016lol} 
  L.~McGough, M.~Mezei and H.~Verlinde,
  %``Moving the CFT into the bulk with $T\bar T$,''
  arXiv:1611.03470 [hep-th].
  %%CITATION = ARXIV:1611.03470;%%
  %3 citations counted in INSPIRE as of 08 Mar 2017
  
    
  \bibitem{polyakov}
A. Polyakov,
``Gauge Fields and Strings", Chapter 9,
Harwood Academic Publishers (1987) ISBN 3-7186-0393-4

  %\cite{Kyono:2017jtc}
\bibitem{Kyono:2017jtc} 
  H.~Kyono, S.~Okumura and K.~Yoshida,
  ``Deformations of the Almheiri-Polchinski model,''
  JHEP {\bf 1703}, 173 (2017)
  doi:10.1007/JHEP03(2017)173
  [arXiv:1701.06340 [hep-th]].
  %%CITATION = doi:10.1007/JHEP03(2017)173;%%
  %3 citations counted in INSPIRE as of 07 Nov 2017

  
  \end{thebibliography}
\end{document}